\documentclass[12pt,a4paper]{article}
\usepackage{amsmath}
\usepackage{amssymb}
\pdfoutput=1
\usepackage[bookmarks=false]{hyperref}
\usepackage{amsthm}
\usepackage{float}
\usepackage{amsfonts}
\usepackage{graphicx}
\usepackage{verbatim}
\usepackage[left=2cm,right=2cm,top=3cm,bottom=2.5cm]{geometry}
\usepackage[numbers]{natbib}
\usepackage[utf8]{inputenc}
\usepackage[usenames,dvipsnames,svgnames]{xcolor}
\usepackage{hyperref}
\hypersetup{colorlinks,allcolors=blue}
\def\myalign#1{
  \def\trule{\noalign{\smallskip\hrule\medskip}}
  \def\nebc{\nearrow\bigcup}
  \def\sebc{\searrow\bigcup}
  \def\pminf{{}_{-\infty}|^{+\infty}}
  \let\Inf\infty
  \def\amp{&} 
  \vbox{\mathsurround0pt\openup1\jot
    \halign{
      &$\displaystyle##\hfil\tabskip0pt$&\amp##\tabskip1em\crcr
      \noalign{\hrule height1pt\smallskip}#1\noalign{\smallskip\hrule height1pt}\crcr}}}    
\begin{document}

\begin{center}
\textbf{Cosmological perturbations in f(G) gravity}
\end{center}

\hfill\\ Albert Munyeshyaka  $^{1}$, Joseph Ntahompagaze $^{2}$, and Tom Mutabazi ${^1}$ \\

\hfill \\
$^{1}$Department of Physics, Mbarara University of Science and Technology, Mbarara, Uganda, \\ $^{2}$ Department of Physics, College of Science and Technology, University of Rwanda, Rwanda 

\hfill \\

Correspondance: munalph@gmail.com\;\;\;\;\;\;\;\;\;\;\;\;\;\;\;\;\;\;\;\;\;\;\;\;\;\;\;\;\;\;\;\;\;\;\;

\begin{center}
\textbf{Abstract}
\end{center}

We explore cosmological perturbations in a modified Gauss-Bonnet f(G) gravity, using a 1+3 covariant formalism. In such a formalism, 
we define gradient variables to get perturbed linear evolution equations.  We transform these linear evolution equations
 into ordinary differential equations using a spherical harmonic decomposition method. The obtained ordinary differential 
equations are time-dependent and then transformed into redshift  dependent. After these transformations, we analyse energy-density 
perturbations for two fluid systems, namely for a Gauss-Bonnet field-dust system and for a Gaus-Bonnet field-radiation system for
three different pedagogical f(G) models: trigonometric, exponential and logarithmic.
For the Gauss-Bonnet field-dust system, energy-density perturbations decay with  increase in redshift for all the three models. 
For the Gauss-Bonnet field-radiation system, the energy-density perturbations decay with increase in redshift for all of the 
three f(G) models for long wavelength modes whereas for short wavelength modes, the energy-density perturbations decay with 
increasing redshift for the logarithmic and exponential f(G) models and oscillate with decreasing amplitude for the trigonometric f(G) model. \\

\hfill \\

\textit{keywards:} $f(G)$ gravity; Covariant formalism; Cosmological perturbations.\\

\section{Introduction}
The discovery of cosmic acceleration \cite{weinberg2013observational, caldwell2009physics,silvestri2009approaches}
motivated enormously the long way to achieve the current standard model in Cosmology. This model called $\Lambda$CDM still 
relies on general relativity with a cosmological constant, to account for such acceleration and, a hypothetical cold dark matter.
The success of this  model relies on its ability
to describe a wide range of different cosmological observations with the most important in this context to be: the spectrum of fluctuations in the cosmic microwave background (CMB), the 
clustering of galaxies, the gravitational lensing observables, the formation and distribution of large scale structures, the Big Bang Nucleosynthesis (BBN)  
\cite{silvestri2009approaches,raveri2019concordance,perlmutter1999measurements,riess1998observational,burles2001big,pogosian1999cosmic,bernardeau2002large}. 
However, the standard model of cosmology $(\Lambda CDM)$ has suffered with the explanation of the cause of 
 the acceleration of the expansion of the universe at least without inclusion of cosmological constant. 
Moreover, there are important open questions in  active  theoretical and observational cosmology such as the horizon problem, the flatness problem, the monopole problem,
the origin and fate of the universe, the formation 
of the primordial universe, the inhomogeneity and anisotropy of the universe 
\cite{debono2016general,li2007cosmology,giblin2016observable,cadoni2020anisotropic},  
 the way cosmological perturbations and primordial fluctuations of the early universe produced the large scale structures, 
and how the astropysical objects such as stars, galaxies were formed and how they have evolved 
\cite{cognola2007string,linde1982new,guth1981inflationary,barreira2016structure,Ananda2009}.\\ \\
The current concordance cosmological 
($\Lambda$ CDM) model \cite{clarkson2010inhomogeneity} assumes that the universe had to go through different epochs such as inflation, 
radiation-dominated, matter-dominated and dark energy-dominated.
The decay of energy-density perturbations with an increase in redshift is the main reason behind the inhomogeneity and large scale 
structure formation of the universe.
There are different literatures for both GR \cite{dunsby1991gauge} and modified gravity theories 
\cite{ellis1999cosmological,abebe2012covariant,kodama1984cosmological,li2018effective,li2007cosmology,sharif2017stability,de2009construction,
boehmer2009stability,nojiri2007introduction} talking about cosmological perturbations.
However, we can study different aspects of cosmology in modified theories of gravity including $f(G)$ gravity.
\cite{li2007cosmology} studied cosmology of modified Gauss-Bonnet gravity and showed how the $f(G)$ models are highly constrained by cosmological data.
 In this work, the study of cosmological perturbations in modified Gauss-Bonnet $f(G)$ gravity using a $1+3$ covariant formalism is the main focus. \\ \\
In this context, we define the gradient variables of Gauss-Bonnet fluid in addition to the gradient variables
of the physical standard matter fluids to derive the perturbation equations.
For further analysis, we use the quasi-static approximation where we compare very slow temporal fluctuations in perturbations of both Gauss-Bonnet 
energy-density and momentum with the fluctuations of matter energy-density, 
as such, we neglect the time derivative terms of the fluctuations of Gauss-Bonnet energy-density and momentum.\\ \\
  Moreover, we consider different pedagogical models namely,\\  
$f_{1}(G)=\frac{\alpha G}{\sqrt{G0}}\arctan(\frac{G}{G0})-\alpha \lambda \sqrt{G0}$ \cite{de2009construction},  
$f_{2}(G)=-M_{pl}^{2}\Lambda (1-\alpha\exp^{-\frac{G}{G0}})$ \cite{inagaki2020gravitational} and $f_{3}(G)=\ln(\alpha \frac{G}{G0})$ 
\cite{zhou2009cosmological} for analysis.
Finally, we analyse the growth of energy-density perturbations with redshift for both $f(G)=G$ and $f(G)$ gravity approaches.
The energy-density perturbations in a Gauss-Bonnet-dust system decay with increase in redshift for all the three models.
  For radiation-dominated universe, the energy-density perturbtions for long wavelength modes decay with increase in redshift for all the three models
whereas the energy-density perturbations for short wavelength modes decay with increase in redshift for all the three models and 
oscillate with a decreasing amplitude for the 
trigonometric model.\\ \\
The roadmap of this paper is as follow: In the following section, we review the $1+3$ covariant formalism within an $f(G)$ gravity 
framework. We derive the linear evolution equations for matter and Gauss-Bonnet perturbations in Section $3$ whereas in Section $4$, we explore second-order 
evolution equations. In Section $5$, we apply the harmonic decomposition method for scalar perturbation equations where partial 
differential equations are reduced to ordinary differential equations.
 Section $6$ is devoted to matter energy-density fluctuations in $f(G)$ gravity. In Section $7$, we apply the redshift transformation method, and we then consider 
 specific pedagogical $f(G)$ models and numerical results in Section $8$. 
In Section $9$, we discuss our results which finally leads to the conclusion in Section $10$. \\ \\
The adopted spacetime signature is $(-,+,+,+)$ and unless stated otherwise, we use $a,b...=0,1,2, 3$ and $8\pi G_{N}=c=1$, where $G_{N}$ is the gravitational constant 
and $c$ is the speed of light and we consider  Friedmann-Robertson-Walker (FRW) space-time background.
\section{The $1+3$ covariant formalism in $f(G)$ gravity theory} 
The main idea behind the $1+3$ covariant formalism is to make spacetime splits of physical quantities with respect to the $4$-velocity $ u^{a}$ of an 
observer.\\ \\
Some authors like  \cite{fujii2003scalar} considered  perturbations of several quantities like the energy-density parameter, expansion parameter and those 
of curvature  
 and  \cite{Ananda2009} applied the $1 + 3$ covariant formalism  to study linear perturbation in general
relativity  and  $f (R)$ gravity.
In $f(G)$ gravity, we study the $1 + 3$ covariant
linear perturbation under FRW background. We divide spacetime into foliated hypersurfaces with constant $G$ (where $G$ is the 
Gauss-Bonnet parameter) and
a perpendicular 4-vector field in the vicinity of $f(G)$ gravity theory. We decompose the cosmological manifold $(M,g)$ 
 into the submanifold $(M,h)$ with a perpendicular 4-velocity field
vector $u^{a}$. The 4-velocity field
vector $u^{ a}$ is defined as
\begin{equation}
 u^{a}=\frac{dx^{a}}{d\tau},
\end{equation}
where $\tau$ is the proper time such that $ u^{a}u_{a}=-1$. The metric $g_{ ab}$ is related to the projection tensor $h_{ ab} $ via: 
 \begin{equation}
 g_{ab} =h_{ab} - u_{a}u_{b}.
 \end{equation}
 Here, the parallel projection tensor is defined as
 \begin{equation}
  U^{a}_{b}=- u_{a}u_{b} \Rightarrow U_{c}^{a}U_{b}^{c}=U_{b}^{a} ,U_{a}^{a}=1,u_{ab}u^{b}=u_{a},
 \end{equation}
 and the orthogonal projection tensor as
 \begin{equation}
  h_{ab}=g_{ab} +u_{a}u_{b} \Rightarrow h^{a}_{c}h_{b}^{c}=h_{b}^{a},h_{a}^{a}=3,h_{ab}u^{b}=0.
 \end{equation}
 The  tensor $u_{b}^{a}$ projects parallel to the 4-velocity vector $ u^{a} $ and the $ h_{ab}$
is responsible for the metric properties of instantaneous restspaces of observers
moving perpendicularly with 4-velocity $u^{a}$. The derivatives also follow with respect
to those projectors. The covariant time derivative (for a given tensor $T_{cd}^{ab}$) is given
as
\begin{equation}
 \dot{T}_{cd}^{ab}=u^{e}\tilde{\bigtriangledown}_{e}T_{cd}^{ab}.
\end{equation}
\\ \\
The action of modified $f(G)$ gravity is given by
\begin{equation}
 S=\frac{1}{2\kappa^{2}}\int d^{4}x \sqrt{-g}(f(G)+ \mathcal{L} ),
\end{equation}
  where $\kappa = 8\pi G_{N}$ is a constant, $f(G)$ is a differentiable function of the Gauss-Bonnet term and $\mathcal{L}$ is lagrangian.
The Gauss-Bonnet term is given by~ 
 \begin{equation}
  G=R^{2}-4R_{\mu \nu}R^{\mu\nu}+R_{\mu\nu\lambda\sigma}R^{\mu\nu\lambda\sigma},
 \end{equation}
where  $R$, $R_{ab}$ and $R_{abcd}$ are the Ricci
scalar, Ricci tensor and Riemann tensor respectively. Varying this action with respect to the metric $g_{ab}$ gives modified Einstein field equations.\\ \\
Einstein field equations of GR provide a specific way in which the metric is
determined from the content of the spacetime. The information about the content
is contained within the energy-momentum tensor $ T_{ab} $. Einstein field equations
of GR relate $ G_{ab} $ and $ T_{ab}$ linearly \cite{pearson2013generalized},
\begin{equation}
 G_{ab}=8\pi G_{N}T_{ab},
\end{equation}
with 
\begin{equation}
 G_{ab}\equiv R_{ab}-\frac{1}{2}g_{ab}R, 
\end{equation}
 the Einstein tensor obtained after combining the Ricci tensor and Ricci scalar.
 In the $1+3$ approach, the kinematic quantities which are obtained from irreducible parts
of the decomposed $\bigtriangledown_{ a}u_{ b}$ are given as \cite{nojiri2007introduction} 
\begin{equation}
 \bigtriangledown_{a}u_{b}=\tilde{\bigtriangledown}  _{a}u_{b}- u_{a}\dot{u_{b}}=\frac{1}{3}\theta h_{ab}+\sigma_{ab}+\omega_{ab}-u_{a}\dot{u_{b}},
\end{equation}
where $\theta \equiv \tilde{\bigtriangledown}  _{a}u^{a}$ is the volume expansion rate of the fluid, with $\theta =3H$, 
$\sigma_{ab}\equiv \tilde{\bigtriangledown}  _\langle{au_{b}}\rangle$ is the symmetric, trace-free rate of shear tensor ($\sigma_{ab}=\sigma_{(ab)}$, ~$\sigma_{ab}u^{b}=0$, ~$\sigma_{a}^{a}=0$)
and describes the rate of distortion of the fluid flow, and $\omega_{ab}\equiv \tilde{\bigtriangledown}[  _{au_{b}}]$ is the 
skew-symmetric vorticity tensor ($\omega_{ab}=\omega[_{ab}]$, ~$\omega_{ab}u^{b}=0$), describing the rotation of the fluid relative
 to a non-rotating frame. The relativistic acceleration vector $a_{a}\equiv \dot{u}_{a}=u_{a;b}u^{b}$ represents the effects of non-gravitational forces (such as pressure)
 and vanishes for a particle moving under gravitational or inertial forces. The representative length scale is the cosmological scale factor $a(\tau)$
 defined in terms of the expansion $\theta$ and the hubble parameter  $H(\tau)$ as 
 \begin{equation}
  \frac{\dot{a}}{a}=\frac{\theta}{3}=H.
 \end{equation}
These are the quantities that tell us about the overall spacetime kinematics, it means the expansion, shear and vorticity of the fundamental worldlines.
\\ \\
The matter energy-momentum tensor $T_{ab}$ is also decomposed with the $1+3$ covariant approach and it is given as  \cite{abebe2015breaking, carloni2010conformal} 
\begin{equation}
 T_{ab}=\rho u_{a}u_{b} +q_{a}u_{b}+u_{a}q_{b} +p h_{ab} +\pi_{ab},
\end{equation}
where $\rho=T_{ab}u^{a}u^{b}$ is the relativistic energy density, $q^{a}=-T_{ab}u^{b}h^{ca}$ is the relativistic momentum density
(energy flux relative to $u ^{a}$ ), $p=\frac{1}{3}(T_{ab}h^{ab})$ is relativistic isotropic pressure and $\pi_{ ab}=T_{cd}h^{c}\langle ah_{b}^{d}\rangle$ is the trace-free anisotropic pressure of the fluid ($\pi^{a}_{ a}  = 0, \pi_{ ab} = \pi_{ (ab)} , \pi_{ab} u^{b} = 0$). 
These are the dynamical quantities obtained from the energy-momentum tensor. The energy–
momentum tensor for the perfect fluid can be recovered by setting ($q_{ a} = \pi_{ ab} = 0$)
and that tensor $T_ {ab}$ leads to
\begin{equation}
 T_{ab} =\rho u_{a}u_{b}+ ph_{ab},
\end{equation}
where the equation of state (EoS) for perfect fluid is $p_{m} = w\rho_{m}$.
The trace of the energy momentum tensor above is given by 
\begin{equation}
 T=T_{a}^{a}=3p-\rho, 
\end{equation} 
which is used in the derivation of evolution equations.\\ \\
In the effective energy-momentum tensor approach, the Einstein field equations preserve their forms, but the dynamical quantities should be 
replaced with the effective total $\rho^{tot}=\rho^{m}+\rho^{G}$, $p^{tot}=p^{m}+p^{G}$ in which a superscript $G$ means the contribution from 
the Gauss-Bonnet correction. We present the modified Einstein field equations in the form
\begin{equation}
 G_{ab}\equiv R_{ab}-\frac{1}{2}g_{ab}R=8\pi G_{N}T^{total}_{ab}=8\pi G_{N}(T^{m}_{ab}+T^{G}_{ab}).
\end{equation}
 We assume that the non-interacting matter fluid ($\rho= \rho_{m}+\rho_{G}$) with Gauss-Bonnet fluid in the entire universe and the 
 growth of the matter energy density perturbations have a significant role for large scale structure formation. We define gradient variables  for matter
  fluids and Gauss-Bonnet fluids  in the next subsection in order to derive evolution equations.
 \subsection{Matter fluids}
 Considering a homogeneous and isotropic expanding (FRW) cosmological background, let us define spatial gradient variables such as those of energy density 
 and the volume expansion of the fluid as 
 \begin{equation}
 X_{a}=\tilde{\bigtriangledown} _{a}\rho_{m} .
\end{equation}
Then from this quantity, we can define the following gauge invariant variable
\begin{equation}
 D^{m} _{a}=\frac{a}{\rho} X_{a}= \frac{a}{\rho}\tilde{\bigtriangledown}   _{a} \rho_{m},
 \label{eq16}
\end{equation}
here $m$ is not a running index, it only specifies matter.
The ratio $\frac{a}{\rho}$ helps to evaluate the magnitude of energy density perturbations relative
to the background.\\ \\
Further, we define another quantity \cite{ellis1999cosmological, ntahompagaze2018study}, 
the spatial gradient
of the volume expansion
\begin{equation}
 Z_{a}=a\tilde{\bigtriangledown}   _{a} \theta .
 \label{eq17}
\end{equation}
These two gradient variables define comoving fractional density gradient and comoving gradient of the expansion respectively and can in principle be measured observationally \cite{abebe2012covariant}. 
\subsection{Gauss-Bonnet fluids}
Analogously to the $1+3$ cosmological perturbations treatment for $f(R)$ gravity theory, let us define extra key variables resulting from spatial 
gradient variables which are connected with the Gauss-Bonnet fluids for $f(G)$.
We define two other gradient variables $\mathcal{G}_{a}$  and $\mathsf{G}_{a}$ that characterize perturbations due to Guass-Bonnet parameter $G$ and its momentum   $\dot{G}$  and describe the 
inhomogeneities in the gauss-Bonnet fluid
\begin{equation}
\mathcal{G}_{a}=a\tilde{\nabla}  _{a}G,
\label{eq18}
\end{equation}
and 
\begin{equation}
 \mathsf{G}_{a}=a\tilde{\nabla}  _{a}\dot{G}.
 \label{eq19}
\end{equation}
\\ \\
All these gradient variables defined in Eq. \ref{eq16} through to Eq. \ref{eq19} shall be considered to develop the system of cosmological perturbation equations for $f(G)$
gravity in the $(1+3)$ covariant formalism. Moreover, for each non-interacting fluid, the following conservation equations for the energy momentum tensor
\begin{equation}
 \dot{\rho}=-\theta(\rho+p),
 \label{eq20}
\end{equation}
\begin{equation}
 \tilde{\bigtriangledown}_{a}p+(\rho+p)\dot{u}_{a}=0,
 \label{eq21}
\end{equation}
hold, and for non-interacting fluids, we consider the equation of state to be 
\begin{equation}
 p=\omega \rho,
 \label{eq22}
\end{equation}
where $\omega$ is the equation of state parameter. We also consider the propagation equation for expansion
\begin{equation}
 \dot{\theta}-\tilde{\bigtriangledown}_{a}\dot{u}^{a} =-\frac{1}{3}\theta-\frac{1}{2}(\rho+3p),
 \label{eq23}
\end{equation}
which is the Raychaudhuri equation \cite{kar2007raychaudhuri} for which it is the basic equation of gravitational attraction.
This equation can be obtained from  the decomposition of the Riemann tensor and make use of Einstein equations.
The term $\rho+3p$ represents the active gravitational mass density. We use the set of equation Eq. \ref{eq16} through
to Eq. \ref{eq23} to derive linear evolution equations in the next subsection.
\section{Linear evolution equations for matter and Gauss-Bonnet fluid perturbations}
\subsection{General equations}
In this section, we derive first order linear evolution equations for the defined gradient variables. In the energy frame of matter fluid, these evolution equations  for the cosmological
perturbations are given as 
\begin{equation}
 \dot{D_{a}^{m}}=-\left(1+w\right)Z_{a}+\theta  w D^{m}_{a}.
 \label{eq24}
\end{equation} 
This equation is also obtained in many works such as \cite{Ananda2009, Sahlu2020} and it depicts how the expansion hinders the growth of the density perturbations.\\ \\
The linear evolution equation for comoving volume expansion  
 \begin{multline}
     \dot{Z}_{a}=-\frac{w}{1+w}\tilde{\bigtriangledown}^{2}D^{m}_{a}-\left( \rho_{m}+\frac{w}{1+w}\dot{\theta} \right) D^{m}_{a}
     +\frac{4\theta^{3}}{9}f''\mathsf{G}_{a}+\lbrace f'' \left(\frac{1}{2}+\theta f'-G \right)\\
     +f''' \left( \frac{2\theta^{3}\dot{G}}{3}-\frac{3G\dot{G}}{2\theta}-\frac{4\theta^{2}\dot{G}}{3}-\frac{2\theta^{2}\ddot{G}}{3} \right)
     +\frac{f'''}{f''}(\frac{9G}{16\theta^{2}}+\frac{\theta^{2}}{6}+\frac{1-3w}{2}\rho_{m}\\
     -f+Gf'-\frac{2\theta^{2}\dot{G}^{2}}{3}f''')-f'-\frac{9}{16\theta^{2}} \rbrace \mathcal{G}_{a}
     +\lbrace \frac{1-3w}{\theta}\rho_{m}-\frac{2\theta}{3}+\frac{9}{4\theta^{2}}-\frac{1}{2\theta}f\\
     +\frac{G}{2\theta}f'-\frac{4\theta\dot{G}^{2}}{3}f'''+f'' \left( \frac{8\theta^{2}\dot{G}}{9}+\frac{3G\dot{G}}{2\theta^{2}}
     -\frac{4\theta\ddot{G}}{3}-\frac{9G\dot{G}}{2\theta^{2}}\right)\rbrace Z_{a},
     \label{eq25}
 \end{multline}
 is obtained by taking into account the trace part of the Einstein's field equations and that of energy momentum tensor.\\ \\
 The derivation of evolution equation for $\mathcal{G}_{a}$  is straightfoward and yields
\begin{equation}
  \dot{\mathcal{G}_{a}}=\mathsf{G}_{a}-\frac{w}{\left(1+w\right)}\dot{G} D^{m}_{a}.
  \label{eq26}
 \end{equation} 
 Finally, we obtain the evolution equations for $\mathsf{G}_{a}$  by  taking the spatial gradient of the trace equation
 \begin{multline}
  \dot{\mathsf{G}_{a}}=\lbrace \frac{3\left(1-3w\right)}{4}\frac{\rho_{m}}{\theta^{2}f''}\rbrace D^{m}_{a}+\lbrace -\frac{\theta}{3}-\frac{9}{4}\frac{G}{\theta^{3}}-2\frac{f'''\dot{G}}{f''}\rbrace \mathsf{G}_{a}+\lbrace \frac{1}{\theta^{2}f''}\left(\frac{27}{32\theta^{2}}+\frac{3}{2}f'\right)\\
  +\frac{f'''}{\theta^{2}f''}\left(-\frac{27}{32}\frac{G}{\theta^{2}}-\frac{\theta^{2}}{4}
  -\frac{3\left(1-3w\right)}{4}\rho_{m}+\frac{3}{2}f-\frac{3}{2}Gf'+\dot{G}^{2}\theta^{2}f'''\right)-\frac{3}{2\theta}f'\\
  +\frac{3}{2\theta^{2}}G-\frac{9}{4\theta^{3}}\dot{G}-\frac{\dot{G}^{2}f^{iv}}{f''}\rbrace \mathcal{G}_{a}
  +\lbrace \frac{1}{\theta^{2}f''}\left(-\frac{27}{8\theta^{3}}-\frac{3\left(1-3w\right)\rho_{m}}{2\theta}+\frac{3}{\theta}f-\frac{3}{\theta}Gf'\right)\\
  -\frac{\dot{G}}{3}+\frac{27}{4\theta^{4}}G\dot{G}\rbrace Z_{a}+a\ddot{G}\dot{u}_{a}.
  \label{eq27}
 \end{multline}
\\ \\
 We get the linear evolution equations (Eq. \ref{eq24} through to Eq. \ref{eq27}) after  differentiating the gradient variables (Eq. \ref{eq16} through to Eq. \ref{eq19}) 
 with respect to cosmic time and make use of  linear covariant identity for any scalar quantity $f$,
\begin{equation}
  (\tilde{\bigtriangledown}_{a}f)\dot{}=\tilde{\bigtriangledown}_{a}\dot{f}- \frac{\theta }{3} \tilde{\bigtriangledown}_{a}f +\dot{f}\dot{u}_{a},
\end{equation} and linealized form of the propagation equations (Eq. \ref{eq20}, Eq. \ref{eq21} and Eq. \ref{eq23}).\\ \\
  Eq. \ref{eq25} through to Eq. \ref{eq27} are new, together with Eq. \ref{eq24}, they describe the evolution of the gradient variables and 
  inhomogeneities in the matter for a general $f(G)$  theory of gravity.
 \subsection{Scalar decomposition}
 The vector gradient variable equations  described are general evolution equations of the perturbations, but only scalar part of the gradient variables  are 
understood to play a key role in matter clustering and hence in structure formation. The linear temporal scalar evolution equations are therefore given as 
\begin{equation}
 \dot{\Delta}^{m}=-(1+w)Z+w\theta \Delta^{m},
 \label{eq29}
\end{equation}
this equation has been obtained in the work done in \cite{Ananda2009} and \cite{Sahlu2020}.
 \begin{multline}
  \dot{Z}= -\frac{w}{1+w} \tilde{\bigtriangledown}^{2}\Delta^{m}-\left(\rho_{m}+\frac{w}{1+w}\dot{\theta}\right)\Delta^{m}+\frac{4\theta^{3}}{9}f''\mathsf{G}+ \{\frac{1}{2}f''+f''\left(\theta f'-G\right)\\
  +f'''\left(\frac{2\theta^{3}\dot{G}}{3}-\frac{3G\dot{G}}{2\theta}-\frac{4\theta^{2}\dot{G}}{3}-\frac{2\theta^{2}\ddot{G}}{3}\right) +\frac{f'''}{f''} ( \frac{9G}{16\theta^{2}}+\frac{\theta^{2}}{6}+\frac{1-3w}{2}\rho_{m}\\
  -f+Gf'-\frac{2\theta^{2}\dot{G}^{2}}{3}f''')-f'-\frac{9}{16\theta^{2}}\}\mathcal{G}+ \{\frac{1-3w}{\theta}\rho_{m}-\frac{2\theta}{3}+\frac{9}{4\theta^{2}}-\frac{1}{2\theta}f\\
  +\frac{G}{2\theta}f'-\frac{4\theta\dot{G}^{2}}{3}f'''+f''\left(\frac{8\theta^{2}\dot{G}}{9}+\frac{3G\dot{G}}{2\theta^{2}}-\frac{4\theta\ddot{G}}{3}-\frac{9G\dot{G}}{2\theta^{2}}\right) \}Z,
  \label{eq30}
 \end{multline}
\begin{equation}
 \dot{ \mathcal{G}}=\mathsf{G}-\frac{w}{1+w}\dot{G}\Delta^{m},
 \label{eq31}
\end{equation}
 and 
  \begin{multline}
 \dot{\mathsf{G}}=\{ \frac{3(1-3w)}{4}\frac{\rho_{m}}{\theta^{2}f''}-\frac{w}{1+w}\ddot{G}\}\Delta^{m}+\{-\frac{\theta}{3}-\frac{9}{4}\frac{G}{\theta^{3}}-2\frac{f'''\dot{G}}{f''}\}\mathsf{G}+\{\frac{1}{\theta^{2}f''}\left(\frac{27}{32\theta^{2}}+\frac{3}{2}f'\right)\\
 +\frac{f'''}{\theta^{2}f''}\left(-\frac{27}{32}\frac{G}{\theta^{2}}-\frac{\theta^{2}}{4}
  -\frac{3\left(1-3w\right)}{4}\rho_{m}+\frac{3}{2}f-\frac{3}{2}Gf'+\dot{G}^{2}\theta^{2}f'''\right)-\frac{3}{2\theta}f'\\
  +\frac{3}{2\theta^{2}}G-\frac{9}{4\theta^{3}}\dot{G}-\frac{\dot{G}^{2}f^{iv}}{f''}\}\mathcal{G}
  +\{\frac{1}{\theta^{2}f''}\left(-\frac{27}{8\theta^{3}}-\frac{3\left(1-3w\right)\rho_{m}}{2\theta}+\frac{3}{\theta}f-\frac{3}{\theta}Gf'\right)\\-\frac{\dot{G}}{3}+\frac{27}{4\theta^{4}}G\dot{G}\}Z.
  \label{eq32}
\end{multline}
Eq. \ref{eq30} through to Eq. \ref{eq32} are new. To get the above linear scalar evolution equations, we extract the scalar part of the vector gradient variables using scalar decomposition method then make their temporal derivatives.
 For further analysis, we derive  second order linear evolution equations in the next subsection.
 \section{Second order linear  evolution equations}
 We obtain a set of second order linear evolution equations from equations Eq. \ref{eq24} through to Eq. \ref{eq32} by making the second derivative of gradient
  variables with respect to cosmic time.
This has  advantage of simplifying the equations and make them manageable. After differentiating Eq. \ref{eq29} and making little algebra, we  have
\begin{multline}
 \ddot{\Delta^{m}}=w \tilde{\bigtriangledown}^{2}\Delta^{m}+ \{\frac{1-3w}{\theta}\rho_{m}
 -\frac{2\theta}{3}+\frac{9}{4\theta^{2}}-\frac{1}{2\theta}f+\frac{G}{2\theta}f'-\frac{4\theta\dot{G}^{2}}{3}f'''\\
 +f''\left(\frac{8\theta^{2}\dot{G}}{9}+\frac{3G\dot{G}}{2\theta^{2}}-\frac{4\theta\ddot{G}}{3}-\frac{9G\dot{G}}{2\theta^{2}} \right)+w\theta\}\dot{\Delta^{m}}
 +\{\left(1+w\right)\rho_{m}+2w\dot{\theta}\\
 + \{\frac{1-3w}{\theta}\rho_{m}-\frac{2\theta}{3}+\frac{9}{4\theta^{2}}-\frac{1}{2\theta}f+\frac{G}{2\theta}f'-\frac{4\theta\dot{G}^{2}}{3}f'''
 +f''(\frac{8\theta^{2}\dot{G}}{9}+\frac{3G\dot{G}}{2\theta^{2}}\\
 -\frac{4\theta\ddot{G}}{3}-\frac{9G\dot{G}}{2\theta^{2}})
 -4w \theta^{2}\dot{G}\}\left(-w\theta \right)\}\Delta^{m}-(1+w)\frac{4\theta^{3}}{9}f''\dot{\mathcal{G}}
 -\left(1+w\right) \{f'' (\frac{1}{2}\\+\theta f'-G)
 +f'''\left(\frac{2\theta^{3}\dot{G}}{3}-\frac{3G\dot{G}}{2\theta}-\frac{4\theta^{2}\dot{G}}{3}-\frac{2\theta^{2}\ddot{G}}{3} \right)
 +\frac{f'''}{f''} (\frac{9G}{16\theta^{2}}+\frac{\theta^{2}}{6}\\+\frac{1-3w}{2}\rho_{m}
 -f+Gf'-\frac{2\theta^{2}\dot{G}^{2}}{3}f''' )-f'-\frac{9}{16\theta^{2}}\}\mathcal{G},
 \label{eq33}
 \end{multline}
 and after differentiating Eq. \ref{eq31} and making little algebra, we have
  \begin{multline}
 \ddot{ \mathcal{G}}=\{\frac{1}{\theta^{2}f''}\left(-\frac{27}{8\theta^{3}}-\frac{3(1-3w)\rho_{m}}{2\theta}+\frac{3}{\theta}f-\frac{3}{\theta}Gf'\right)-\frac{\dot{G}}{3}+\frac{27}{4\theta^{4}}G\dot{G}\\
 +w\dot{G}\}(-\frac{1}{1+w})\dot{\Delta^{m}}
 +\{ \frac{3(1-3w)}{4}\frac{\rho_{m}}{\theta^{2}f''}-\frac{2w}{1+w}\ddot{G}+\{\frac{1}{\theta^{2}f''}(-\frac{27}{8\theta^{3}}\\
 -\frac{3(1-3w)\rho_{m}}{2\theta}+\frac{3}{\theta}f-\frac{3}{\theta}Gf')
 -\frac{2\dot{G}}{3}+\frac{9}{2\theta^{4}}G\dot{G}-2\frac{(\dot{G})^{2}}{\theta}\frac{f'''}{f''}\}\left(\frac{w\theta}{1+w}\right)\}\Delta^{m}\\
 +\{-\frac{\theta}{3}-\frac{9}{4}\frac{G}{\theta^{3}}-2\frac{f'''\dot{G}}{f''}\}\dot{\mathcal{G}} 
 +\{\frac{1}{\theta^{2}f''}\left(\frac{27}{32\theta^{2}}+\frac{3}{2}f'\right)
 +\frac{f'''}{\theta^{2}f''}(-\frac{27}{32}\frac{G}{\theta^{2}} -\frac{\theta^{2}}{4}\\
  -\frac{3(1-3w)}{4}\rho_{m}+\frac{3}{2}f-\frac{3}{2}Gf'+\dot{G}^{2}\theta^{2}f''')
  -\frac{3}{2\theta}f'+\frac{3}{2\theta^{2}}G-\frac{9}{4\theta^{3}}\dot{G} 
  -\frac{\dot{G}^{2}f^{iv}}{f''}\}\mathcal{G}.
  \label{eq34}
  \end{multline}
Eq. \ref{eq33} and Eq. \ref{eq34} are new. These are scalar gradient variables ( Eq. \ref{eq29} through to Eq. \ref{eq34}) we take as input to study the energy density fluctuations in different 
 fluid systems namely Gauss-Bonnet field-dust system
  and Gauss-Bonnet field-radiation system after applying the harmonic decomposition method
 of these variables in the next section.
 \section{Spherical harmonic decomposition}
The spherical harmonic decomposition approach is used to get eigenfunctions with the corresponding wavenumber for a harmonic oscillator differential 
equation after applying separation of variables to that second order differential equation. The above evolution equations can be taken as 
a coupled system of harmonic oscillator of the form  \cite{abebe2015breaking}
\begin{equation}
 \ddot{X}+A_{1}\dot{X}+A_{2}X=A_{3}(Y,\dot{Y}),
\end{equation}
where $A_{1}$, $A_{2}$ and $A_{3}$ represent damping (friction) term, the restoring force term and the source forcing term respectively.
A key assumption in the analysis of the equation here is that we can apply the separation  of variables technique such that  
\begin{equation}
 X=\sum_{k} X^{k}(t).Q_{k}(x),
\end{equation}
and 
\begin{equation}
 Y=\sum_{k}Y^{k}(t).Q_{k}(x),
\end{equation} 
where $Q_{k}$ are the eigenfunctions of the covariant Laplace-Beltrami operator such that
 \begin{equation}
  \tilde{\bigtriangledown}^{2}Q=-\frac{k^{2}}{a^{2}} Q^{k},
 \end{equation}
 and  the order of harmonic (wave number) $k$ is given as
 \begin{equation} 
 k=\frac{2\pi a}{\lambda},
\end{equation}
where  $\lambda$ is the physical wavelength of the mode.
This equation represents the relationship between the wavenumber to a cosmological scale. 
The eigenfunctions $Q$ are time-independent, that means $\dot{Q(x)}=0$.\\ \\
This method has been extensively used for $1+3$ covariant linear perturbations , for example  in the works done in \cite{abebe2015breaking, carloni2006gauge, chakraborty2017gravity}.
In this way, the evolution equations can be converted into ordinary differential equations at each mode $k$ separately.
Therefore, the analysis becomes more easier when dealing with ordinary differential equations rather than the partial differential
equations. After harmonic decomposition, Eq. \ref{eq29} through to Eq. \ref{eq34} can be rewritten in the following form
\begin{equation}
 \dot{\Delta_{m}^{k}}=-(1+w)Z^{k}+w\theta \Delta_{m}^{k},
 \label{eq40}
  \end{equation}
  (the same result can be obtained in the works done in \cite{Ananda2009, Sahlu2020}.)
  \begin{multline}
 \dot{Z^{k}}= \left(\frac{w}{1+w}\frac{k^{2}}{a^{2}}-\rho_{m}+\frac{w}{1+w}\dot{\theta}\right)\Delta_{m}^{k}+\frac{4\theta^{3}}{9}f''\mathsf{G}^{k}+ \{f''\left(\frac{1}{2}+\theta f'-G\right)\\
 +f'''\left(\frac{2\theta^{3}\dot{G}}{3}-\frac{3G\dot{G}}{2\theta}-\frac{4\theta^{2}\dot{G}}{3}-\frac{2\theta^{2}\ddot{G}}{3}\right)
 +\frac{f'''}{f''}(\frac{9G}{16\theta^{2}}+\frac{\theta^{2}}{6}+\frac{1-3w}{2}\rho_{m}-f\\+Gf'
 -\frac{2\theta^{2}\dot{G}^{2}}{3}f''')-f'-\frac{9}{16\theta^{2}}\}\mathcal{G}^{k}+ \{\frac{1-3w}{\theta}\rho_{m}-\frac{2\theta}{3}+\frac{9}{4\theta^{2}}-\frac{1}{2\theta}f
 \\+\frac{G}{2\theta}f'
 -\frac{4\theta\dot{G}^{2}}{3}f'''+f''\left(\frac{8\theta^{2}\dot{G}}{9}+\frac{3G\dot{G}}{2\theta^{2}}-\frac{4\theta\ddot{G}}{3}-\frac{9G\dot{G}}{2\theta^{2}}\right)\}Z^{k},
 \label{eq41}
\end{multline}
\begin{equation}
 \dot{ \mathcal{G}^{k}}=\mathsf{G}^{k}-\frac{w}{1+w}\dot{G}\Delta_{m}^{k},
 \label{eq42}
\end{equation}
\begin{multline}
 \dot{\mathsf{G}^{k}}=\{ \frac{3(1-3w)}{4}\frac{\rho_{m}}{\theta^{2}f''}-\frac{w}{1+w}\ddot{G}\}\Delta_{m}^{k}+\{-\frac{\theta}{3}-\frac{9}{4}\frac{G}{\theta^{3}}-2\frac{f'''\dot{G}}{f''}\}\mathsf{G}^{k}+\{\frac{1}{\theta^{2}f''}\left(\frac{27}{32\theta^{2}}+\frac{3}{2}f'\right)\\
 +\frac{f'''}{\theta^{2}f''}\left(-\frac{27}{32}\frac{G}{\theta^{2}}-\frac{\theta^{2}}{4}
  -\frac{3(1-3w)}{4}\rho_{m}+\frac{3}{2}f-\frac{3}{2}Gf'+\dot{G}^{2}\theta^{2}f'''\right)-\frac{3}{2\theta}f'\\+\frac{3}{2\theta^{2}}G
  -\frac{9}{4\theta}^{3}\dot{G}-\frac{\dot{G}^{2}f^{iv}}{f''}\}\mathcal{G}^{k}
  +\{\frac{1}{\theta^{2}f''}\left(-\frac{27}{8\theta^{3}}-\frac{3(1-3w)\rho_{m}}{2\theta}+\frac{3}{\theta}f-\frac{3}{\theta}Gf'\right)\\-\frac{\dot{G}}{3}+\frac{27}{4\theta^{4}}G\dot{G}\}Z^{k},
  \label{eq43}
\end{multline}
\begin{multline}
 \ddot{\Delta^{k}_{m}}= \{\frac{1-3w}{\theta}\rho_{m}+f''\left(\frac{8\theta^{2}\dot{G}}{9}+\frac{3G\dot{G}}{2\theta^{2}}
 -\frac{4\theta\ddot{G}}{3}
 -\frac{9G\dot{G}}{2\theta^{2}}\right)-\frac{2\theta}{3}+\frac{9}{4\theta^{2}}-\frac{1}{2\theta}f
 +\frac{G}{2\theta}f'\\-\frac{4\theta\dot{G}^{2}}{3}f'''+w\theta\}\dot{\Delta^{k}_{m}}+\{(1+w)\rho_{m}+2w\dot{\theta}+ \{\frac{1-3w}{\theta}\rho_{m}-\frac{2\theta}{3}+\frac{9}{4\theta^{2}}\\
 -\frac{1}{2\theta}f+\frac{G}{2\theta}f'-\frac{4\theta\dot{G}^{2}}{3}f'''+f''\left(\frac{8\theta^{2}\dot{G}}{9}+\frac{3G\dot{G}}{2\theta^{2}}-\frac{4\theta\ddot{G}}{3}-\frac{9G\dot{G}}{2\theta^{2}}\right)-4w 
 \theta^{2}\dot{G}\}\left(-w\theta \right)-w\frac{k^{2}}{a^{2}}\}\Delta^{k}_{m}\\-(1+w)\frac{4\theta^{3}}{9}f''\dot{\mathcal{G}^{k}}
 -(1+w) \{f''\left(\frac{1}{2}+\theta f'-G\right)+f'''\left(\frac{2\theta^{3}\dot{G}}{3}
 -\frac{3G\dot{G}}{2\theta}-\frac{4\theta^{2}\dot{G}}{3}-\frac{2\theta^{2}\ddot{G}}{3}\right)\\
 +\frac{f'''}{f''}\left(\frac{9G}{16\theta^{2}}+\frac{\theta^{2}}{6}+\frac{1-3w}{2}\rho_{m}-f+Gf'-\frac{2\theta^{2}\dot{G}^{2}}{3}f'''\right)-f'-\frac{9}{16\theta^{2}}\}\mathcal{G}^{k},
 \label{eq44}
 \end{multline}
and
\begin{multline}
 \ddot{ \mathcal{G}^{k}}=\{\frac{1}{\theta^{2}f''}\left(-\frac{27}{8\theta^{3}}-\frac{3(1-3w)\rho_{m}}{2\theta}+\frac{3}{\theta}f-\frac{3}{\theta}Gf'\right)-\frac{\dot{G}}{3}+\frac{27}{4\theta^{4}}G\dot{G}+w\dot{G}\}\left(-\frac{1}{1+w}\right)\dot{\Delta^{k}_{m}}\\
 +\{ \frac{3(1-3w)}{4}\frac{\rho_{m}}{\theta^{2}f''}-\frac{2w}{1+w}\ddot{G} +\{\frac{1}{\theta^{2}f''}\left(-\frac{27}{8\theta^{3}}-\frac{3(1-3w)\rho_{m}}{2\theta}+\frac{3}{\theta}f
 -\frac{3}{\theta}Gf'\right)\\
 -\frac{2\dot{G}}{3}+\frac{9}{2\theta^{4}}G\dot{G} -2\frac{(\dot{G})^{2}}{\theta}\frac{f'''}{f''}\}\left(\frac{w\theta}{1+w}\right)\}\Delta^{k}_{m}+\{-\frac{\theta}{3}-\frac{9}{4}\frac{G}{\theta^{3}}-2\frac{f'''\dot{G}}{f''}\}\dot{\mathcal{G}^{k}}\\
 +\{\frac{1}{\theta^{2}f''}\left(\frac{27}{32\theta^{2}}+\frac{3}{2}f'\right) 
 +\frac{f'''}{\theta^{2}f''}\left(-\frac{27}{32}\frac{G}{\theta^{2}} -\frac{\theta^{2}}{4}
  -\frac{3(1-3w)}{4}\rho_{m}+\frac{3}{2}f-\frac{3}{2}Gf'
  +\dot{G}^{2}\theta^{2}f'''\right)\\-\frac{3}{2\theta}f'+\frac{3}{2\theta^{2}}G-\frac{9}{4\theta^{3}}\dot{G} 
  -\frac{\dot{G}^{2}f^{iv}}{f''}\}\mathcal{G}^{k}.
  \label{eq45}
  \end{multline}
  Eq. \ref{eq40} through to Eq. \ref{eq45} couple in one equation in $\Delta^{(k)}_{m}$ for simplicity and be able to make further analysis.
\section{Matter density fluctuations in $f(G)$ gravity}
During the evolution of the universe, there should be an epoch where the Gauss-Bonnet field was dominating over the dustlike in the FRW spacetime cosmology.
  In such an epoch, matter fluctuations can be studied. Futhermore, we assume that those two fluids are non-interacting, 
  and using quasi-static approximation, the time fluctuations in perturbations of the Gauss-Bonnet energy density $\mathcal{G}^k$ and momentum $\mathsf{G}^k$ are assumed to be 
  constant with time which means $\dot{ \mathcal{G}}^k=0$, ~$\dot{ \mathsf{G}}^k=0$ and $\ddot{ \mathcal{G}^{k}}=0$ and with the support that
  Gauss-Bonnet field is part of the background, hence its perturbations have no much interest from the homogeneous universe.
  With these assumptions, from Eq. \ref{eq40} through to Eq. \ref{eq45},  we have
  \begin{multline}
 \ddot{\Delta_{m}^{k}}=\{-w\frac{k^{2}}{a^{2}}+(1+w)\rho_{m}-\frac{4\theta^{3}}{9} wf''\dot{G}
 -\{\frac{1-3w}{\theta}\rho_{m}-\frac{2\theta}{3}+\frac{9}{4\theta^{2}}-\frac{1}{2\theta}f
\\ +\frac{G}{2\theta}f'
 -\frac{4\theta\dot{G}^{2}}{3}f'''
 +f''\left(\frac{8\theta^{2}\dot{G}}{9}+\frac{3G\dot{G}}{2\theta^{2}}-\frac{4\theta\ddot{G}}{3}-\frac{9G\dot{G}}{2\theta^{2}}\right)\}(
 w\Theta) 
 \\ -\{f''\left(\frac{1}{2}+\theta f'-G\right)+f'''\left(\frac{2\theta^{3}\dot{G}}{3}-\frac{3G\dot{G}}{2\theta}-\frac{4\theta^{2}\dot{G}}{3}-\frac{2\theta^{2}\ddot{G}}{3}\right)\\
 +\frac{f'''}{f''}\left(\frac{9G}{16\theta^{2}}+\frac{\theta^{2}}{6}+\frac{1-3w}{2}\rho_{m}-f+Gf'-\frac{2\theta^{2}\dot{G}^{2}}{3}f'''\right)-f'-\frac{9}{16\theta^{2}}\}\\
 \{\frac{1}{A}(1+w)\{ \frac{3(1-3w)}{4}\frac{\rho_{m}}{\theta^{2}f''}
  +\{-\frac{\theta}{3}-\frac{9}{4}\frac{G}{\theta^{3}}-2\frac{f'''\dot{G}}{f''}-\ddot{G}\}w\dot{G}\}\\ 
  -\{f''\left(\frac{1}{2}+\theta f'-G\right)+f'''\left(\frac{2\theta^{3}\dot{G}}{3}-\frac{3G\dot{G}}{2\theta}-\frac{4\theta^{2}\dot{G}}{3}-\frac{2\theta^{2}\ddot{G}}{3}\right)\\
 +\frac{f'''}{f''}\left(\frac{9G}{16\theta^{2}}+\frac{\theta^{2}}{6}+\frac{1-3w}{2}\rho_{m}-f+Gf'-\frac{2\theta^{2}\dot{G}^{2}}{3}f'''\right)-f'-\frac{9}{16\theta^{2}}\}\\
  \{\frac{1}{A}\{\frac{1}{\theta^{2}f''}\left(-\frac{27}{8\theta^{3}}-\frac{3(1-3w)\rho_{m}}{2\theta}+\frac{3}{\theta}f-\frac{3}{\theta}Gf'\right)
  -\frac{\dot{G}}{3}+\frac{27}{4\theta^{4}}G\dot{G}\}(w\theta)\}\Delta_{m}^{k}\\ 
 +\{\frac{1-3w}{\theta}\rho_{m}-\frac{2\theta}{3}+\frac{9}{4\theta^{2}}-\frac{1}{2\theta}f
 +\frac{G}{2\theta}f'-\frac{4\theta\dot{G}^{2}}{3}f'''\\
 +f''\left(\frac{8\theta^{2}\dot{G}}{9}+\frac{3G\dot{G}}{2\theta^{2}}-\frac{4\theta\ddot{G}}{3}-\frac{9G\dot{G}}{2\theta^{2}}\right)+w\theta +\{f''\left(\frac{1}{2}+\theta f'-G\right)+f'''(\frac{2\theta^{3}\dot{G}}{3}-\frac{3G\dot{G}}{2\theta}\\-\frac{4\theta^{2}\dot{G}}{3}-\frac{2\theta^{2}\ddot{G}}{3})
 +\frac{f'''}{f''}\left(\frac{9G}{16\theta^{2}}+\frac{\theta^{2}}{6}+\frac{1-3w}{2}\rho_{m}-f+Gf'-\frac{2\theta^{2}\dot{G}^{2}}{3}f'''\right)-f'-\frac{9}{16\theta^{2}}\}\\
  \{\frac{1}{A}\{\frac{1}{\theta^{2}f''}\left(-\frac{27}{8\theta^{3}}-\frac{3(1-3w)\rho_{m}}{2\theta}+\frac{3}{\theta}f-\frac{3}{\theta}Gf'\right)
  -\frac{\dot{G}}{3}+\frac{27}{4\theta^{4}}G\dot{G}\}\}\dot{\Delta_{m}^{k}},
  \label{eq46}
 \end{multline}
 where $A$ is given in the appendix.
 Eq. \ref{eq46} is new and is the one to be used for energy density perturbations analysis. For 
 the scale factor $a$ of a power law form
 \begin{equation*}
  a=a_{0}\left(\frac{t}{t_{0}}\right)^{\frac{2m}{3(1+w)}}, 
 \end{equation*} one has the volume expansion $\theta$,  
 matter energy density $\rho_{m}$  and the Gauss-Bonnet invariant respectively as
\begin{equation}
 \theta =\frac{2m}{(1+w)t},
 \label{eq47}
\end{equation}
\begin{equation}
{\displaystyle{
 \rho_{m}=\left(\frac{3}{4}\right)^{1-m}\left[\frac{4m^{2}-3m(1+w)}{(1+w)^{2}t^{2}}\right]^{m-1}\frac{4m^{3}-2m(m-1)\left(2m(3w+5)-3(1+w)\right)}{(1+w)^{2}t^{2}}}},
 \label{eq48}
\end{equation}
and \begin{equation}
{\displaystyle{
 G=24\left(\frac{\dot{a}}{a}\right)^{2}\left(\frac{\ddot{a}\dot{a}}{a^{2}}+(\frac{\dot{a}}{a})^{2}\right)}}.
\end{equation}
 In most cases, ~$a_{0}$ and $t_{0}$ are normalized
to unity \cite{abebe2012covariant}. These solutions have been obtained in the work of \cite{carloni2005cosmological}. 
 We make analysis of energy density perturbations for different epochs of the universe namely Gauss-Bonnet field-dust system 
 and Gauss-Bonnet field- radiation system. 
\subsection{ Perturbations in Gauss-Bonnet field-dust dominated universe }
Assuming that the universe is dominated by a Gauss-Bonnet fluid and  dustlike ($w=0$) mixture, the energy density perturbations of radiation 
matter contribution  become negligible. In such a system, energy density perturbations evolve as
 \begin{equation}
 \ddot{\Delta^{k}_{d}}-B_{1}\dot{\Delta_{d}^{k}}-B_{2}\Delta_{d}^{k}=0,
   \label{eq49}
 \end{equation}
 where $B_{1}$ and $B_{2}$ are given in the appendix.
 Here we use $w=0$ in Eq. \ref{eq46} and considered that $\Delta_{m}=\Delta_{d}$ and $\rho_{m}=\rho_{d}$. For the case $f(G)=G$, Eq. \ref{eq49} reduces to
\begin{equation}
 \ddot{\Delta_{d}^{k}}-\frac{2}{3}\theta \dot{\Delta_{d}^{k}}-\frac{1}{2}\rho_{d}\Delta^{k}_{d}=0.
 \label{eq50}
\end{equation}  
\subsection{ Perturbations in Gauss-Bonnet field-radiation dominated universe }
In this part, we assume that the universe was dominated by a Gauss-Bonnet fluid and  radiation mixture as a background, 
where an equation of state parameter is given by  $w=\frac{1}{3}$.
 This results in negligible energy density perturbations of dust matter contribution. In such a system, perturbations would evolve 
according to the following equation ( see Eq. \ref{eq46})
\begin{equation}
 \ddot{\Delta_{r}^{k}}-C\dot{\Delta_{r}^{k}}-D\Delta_{r}^{k}=0,
     \label{eq51}
 \end{equation}
 where $C$ and $D$ are given  in  the appendix.
 Here we insert $w=\frac{1}{3}$  in Eq. \ref{eq46} and used $\Delta_{m}=\Delta_{r}$ and $\rho_{m}=\rho_{r}$.
 For the case $f(G)=G$, Eq. \ref{eq51} 
 reduces to 
 \begin{equation}
 \ddot{\Delta_{r}^{k}}-\theta \dot{\Delta_{r}^{k}}-\{-\frac{k^{2}}{3a^{2}}+\frac{1}{3}\dot{\theta}-\frac{2}{9}\theta^{2}+\frac{4}{3}\rho_{r}\}\Delta^{k}_{r}=0,
 \label{eq52}
\end{equation}
We need the redshift dependent equations so that the analysis of matter energy density perturbations with redshift
be possible.
\section{Redshift transformation}
The scale factor is related to the cosmological redshift as \cite{Sahlu2020}
\begin{equation}
 a=\frac{1}{1+z}.
\end{equation}
For convinience, we also transform any time derivative functions $f$ and $H$ into a redshift derivative as follows:
\begin{equation}
 \frac{\dot{f}}{H}=\frac{df}{dN},
\end{equation}
where $dN\equiv \ln(a)$
\begin{equation}
 \dot{f}=-(1+z)Hf',
\end{equation}
\begin{equation}
 \ddot{f}=(1+z)^{2}H(H'f'+Hf'').
\end{equation}
With little algebra, we have the redshift transformation of the volume expansion $\theta$, Hubble parameter $H$ and Gauss-Bonnet parameter $G$ as 
\begin{equation}
 \theta =\frac{2m}{1+w}(1+z)^{\frac{3(1+w)}{2m}},
 \label{eq59}
\end{equation}
\begin{equation}
 H=\frac{2m}{3(1+w)}(1+z)^{\frac{3(1+w)}{2}},
\end{equation}
and
\begin{equation}
 G=\frac{128}{27}(\frac{m}{1+w})^{4}(1+z)^{\frac{6(1+w)}{m}}(1-(1+z)^{\frac{3(1+z)}{2m}}).
 \label{eq61}
\end{equation}
We use redshift parameter to compare the cosmological behavior of the models with cosmological observations.\\ \\
 After obtaining the final expressions of the energy-density perturbations, we have considered different types of $f(G)$ models: Trigonometric, Exponential 
  and  Logarithmic for a quantitative analysis of the evolution of cosmological perturbations in $f(G)$ gravity. Some of the motivations behind the choice of these models include:
  \begin{itemize}
   \item They are proven to be viable models that are compatible with cosmological observations. \cite{de2009construction}
   \item They are proven to be representative examples of models that could account for the late-time acceleration of the universe without
   the need for dark energy. \cite{li2011large}
 \end{itemize}
\section{Specific pedagogical $f(G)$ models and numerical solutions}
\subsection{The trigonometric $f(G)$ model}
\begin{figure}[ht!]
 \includegraphics[width=120mm,height=100mm]{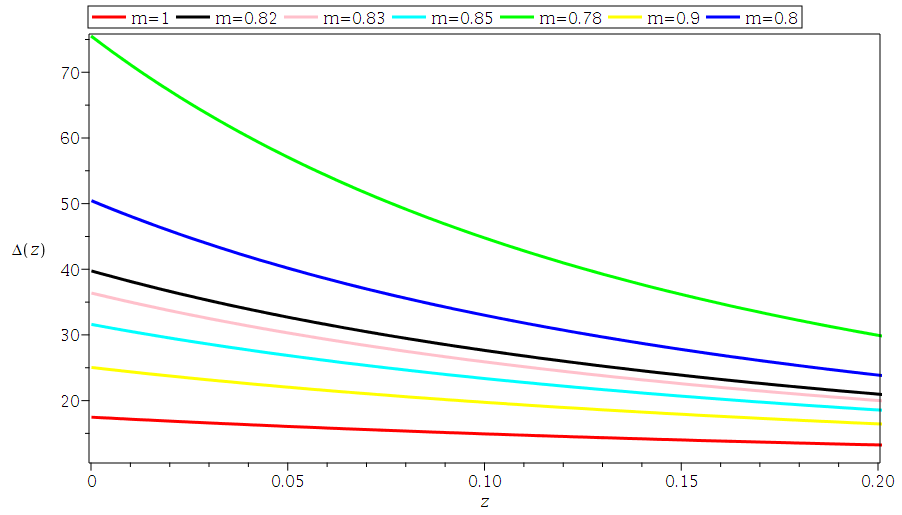}
 \caption[ Plot of energy-density perturbations $\Delta(z)$ as a function of redshift $z$ for $f_{1}(G)$ model, in a dust dominated universe.]
 { Plot of energy-density perturbations $(\Delta(z))$ as a function of redshift $(z)$ for Gauss-Bonnet field-dust system 
 for Eq. \ref{eq49}  and Eq. \ref{eq50}. 
The red line corresponds to the case $f(G)=G$. The assumption 
that $\Delta^{(k)}_{(d)0}=10^{-5}$ and $\Delta'^{(k)}_{(d)0}=0$ was made and the $\Delta(z)$  have been normalized so that the curves coincide on large scale.
As expected 
$\Delta (z)$ decays with increase in redshift. The choice of the values of $m$  follows the work done in \cite{Ananda2009, abebe2013large} throughout this work.}
\label{Fig(1)}
\end{figure}

We consider \cite{de2009construction}
\begin{equation}
f_{1}(G)=\frac{\alpha G}{\sqrt{G0}}\arctan(\frac{G}{G0})-\alpha \lambda \sqrt{G0},
\end{equation}
 for the case $\lambda=0$, $\alpha =1$ and $\frac{1}{\sqrt{G0}}\arctan(\frac{G}{G0})\rightarrow 1$, then $f(G)=G$ as in Eq. \ref{eq50}.\\ \\ 
 By considering that the universe is dominated by the mixture of Gauss-Bonnet field-dust fluids ($w=0$), Eq. \ref{eq48}, Eq. \ref{eq59}  and Eq. \ref{eq61} reduce to 
\begin{equation}
{\displaystyle{
 \rho_{d}=\left(\frac{3}{4}\right)^{1-m}\left((4m^{2}-3m)(1+z)^{-\frac{3}{2m}}\right)^{m-1}\left(\frac{-16m^{3}+26m^{2}-6m}{3}\right)(1+z)^{-\frac{3}{m}}}},
 \label{eq63}
 \end{equation}
 \begin{equation}
  \theta=2m(1+z)^{\frac{3}{2m}},
  \label{eq64}
 \end{equation}
and
\begin{equation}
 {\displaystyle{
  G=\frac{128}{27}m^{4}(1+z)^{\frac{6}{m}}\left(1-(1+z)^{\frac{3}{2m}}\right)}},
  \label{eq65}
 \end{equation}
 respectively.\\ \\ Using the redshift transformation scheme and inserting Eq. \ref{eq63} through to Eq. \ref{eq65} into Eq. \ref{eq49} and
 Eq. \ref{eq50} and considering our $f_{1}(G)\equiv f$,  numerical solutions are found and presented  in Figure \ref{Fig(1)}
\newpage
By considering that the universe is dominated by Gauss-Bonnet field-radiation fluids ($w=\frac{1}{3}$), Eq. \ref{eq48}, Eq. \ref{eq59} and Eq. \ref{eq61} reduce to
\begin{equation}
{\displaystyle{
 \rho_{r}=\left(\frac{3}{4}\right)^{(2-m)}\left(\frac{9m(m-1)}{4}(1+z)^{-\frac{2}{m}}\right)^{m-1}\left(-5m^{3}+8m^{2}-2m\right)(1+z)^{-\frac{4}{m}}}},
 \label{eq66}
\end{equation}
\begin{equation}
  \theta=\frac{3m}{2}(1+z)^{\frac{2}{m}},
  \label{eq67}
 \end{equation}
 and
 \begin{equation}
  G=\frac{3}{2}m^{4}(1+z)^{\frac{8}{m}}(1-(1+z)^{\frac{2}{m}}),
  \label{eq68}
 \end{equation}
 respectively.\\ \\
 At this stage, we can consider the dependency of the wavenumber $k$.
 \subsubsection{Short wavelength solutions}
\begin{figure}[ht!]
 \includegraphics[width=120mm,height=100mm]{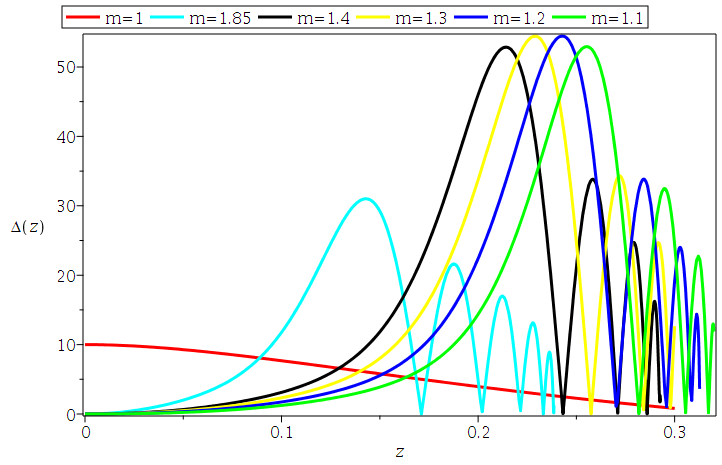}
 \caption[ Plot of energy-density perturbations $(\Delta(z))$ as a function of redshift $(z)$  
 for $f_{1}(G)$ model in a radiation dominated universe, short wavelength modes.]
 {Plot of energy-density perturbations $(\Delta(z))$ as a function of redshift $(z)$ for Eq. \ref{eq51} and Eq. \ref{eq52}  for 
 radiation dominated universe, short-wavelength limit. The red line corresponds to the case where $f(G)=G$. The assumptions 
that $\Delta^{(k)}_{(r)0}=10^{-7}$ and $\Delta'^{(k)}_{(r)0}=0$ were made and using $\frac{k^{2}}{a^{2}H^{2}}\simeq\frac{16\pi^{2}}{3\lambda^{2}(1+z)^4}$, the 
$\Delta(z)$ decays with incease in redshift and shows oscillations features.}
\label{Fig(2)}
\end{figure}

 Here we discuss the growth of energy density fluctuations within the horizon, where $\frac{k^{2}}{a^{2}}\gg 1$. In this regime, the Jeans wavelength $\lambda_{j}$
is much larger than the wavelength of the mean free path of the photon $\lambda_{p}$ and the wavelength of the non interacting fluid, 
it means, $\lambda \ll\lambda_{p} \ll\lambda_{j}$.  Similar analysis was done in \cite{dunsby1991gauge} for GR and \cite{abebe2012covariant} 
for $f(R)$ gravity theory. \\ \\ By knowing that $\frac{k^{2}}{a^{2}H^{2}}\simeq\frac{16\pi^{2}}{3\lambda^{2}(1+z)^4}$ and using the
redshift transformation scheme and inserting Eq. \ref{eq66} through to Eq. \ref{eq68} into Eq. \ref{eq51} and Eq. \ref{eq52} and
using $f_{1}(G)\equiv f$, numerical solutions for Gauss-Bonnet field-radiation dominated system short wavelength modes are found and 
presented in Figure \ref{Fig(2)}.
\newpage
\subsubsection{ Long-wavelength solutions} 
\begin{figure}[h!]
 \includegraphics[width=120mm,height=100mm]{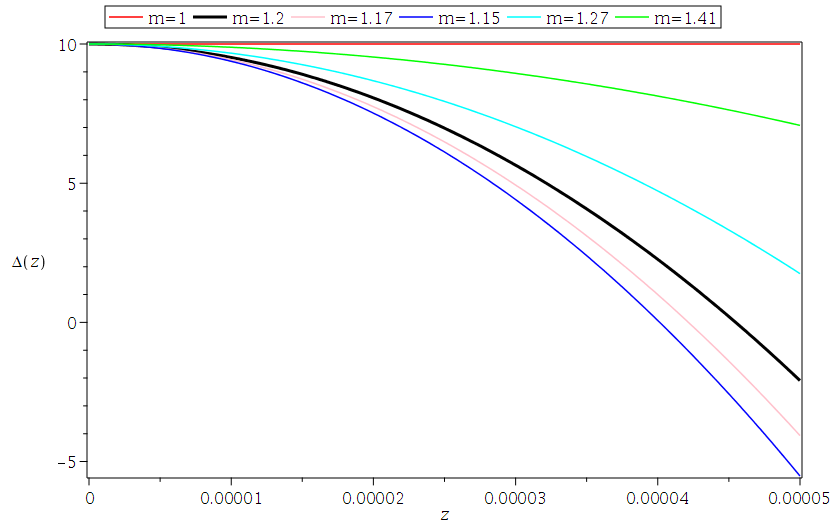}
 \caption[Plot of energy density perturbations $(\Delta(z))$ versus redshift $(z)$  
 for $f_{1}(G)$ model in a 
 radiation dominated universe, long wavelength modes.]
 { Plot of energy-density perturbations $(\Delta(z))$ versus redshift $(z)$ of Eq. \ref{eq51} and  Eq. \ref{eq52} for $f_{1}(G)$ model
in a  radiation dominated universe, long-wavelength limit. The red line corresponds to $f(G)=G$. The assumption 
that $\Delta^{(k)}_{(r)0}=10^{-9}$ and $\Delta'^{(k)}_{(r)0}=0$  was made and we used $\frac{k^{2}}{a^{2}H^{2}}\simeq\frac{16\pi^{2}}{3\lambda^{2}(1+z)^4}\simeq0$.
The $\Delta(z)$ decays with increase in redshift }
\label{Fig(3)}
\end{figure}

The growth of energy density fluctuations is studied for the long-wavelength limit, where $\frac{k^{2}}{a^{2}}\ll 1$ . All cosmological 
fluctuations begin and remain inside the Hubble horizon. With the $k$- dependency dropped out ($\frac{k^{2}}{a^{2}H^{2}}\simeq\frac{16\pi^{2}}{3\lambda^{2}(1+z)^4}\simeq0$) 
and using the redshift transformation scheme and inserting Eq. \ref{eq66}
   through to Eq. \ref{eq68} into Eq. \ref{eq51} and Eq. \ref{eq52} and using $f_{1}(G)\equiv f$, numerical solutions for Gauss-Bonnet field-radiation dominated system long wavelength modes are 
   found and presented in Figure \ref{Fig(3)}.

\newpage
\subsection{The exponential $f(G)$ model}
\begin{figure}[ht!]
 \includegraphics[width=120mm,height=100mm]{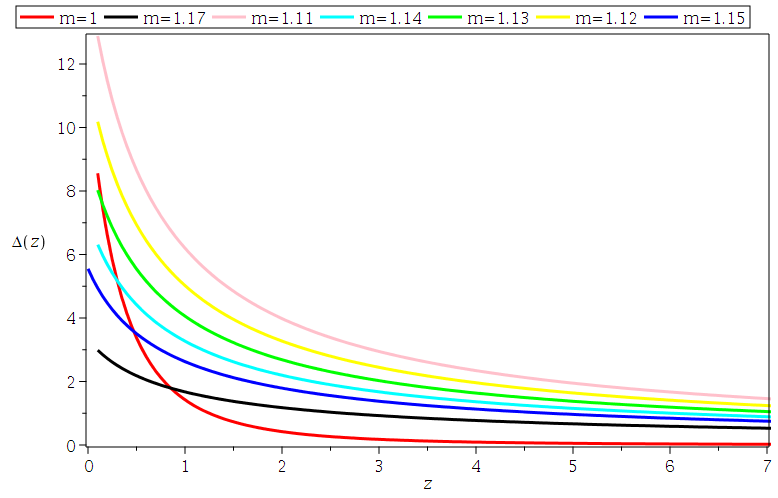}
 \caption[
  Plot of energy-density perturbations $(\Delta(z))$ versus redshift $(z)$ of $f_{2}(G)$ model in a dust dominated universe.]
 {Plot of energy-density perturbations $(\Delta(z))$ versus redshift $(z)$ of Eq. \ref{eq49} and Eq. \ref{eq50}
 for 
$f_{2}(G)$ model in a dust dominated universe.  
The red line corresponds to $f(G)=G$. The assumptions 
that $\Delta^{(k)}_{(d)0}=10^{-5}$ and $\Delta'^{(k)}_{(d)0}=0$ were made and 
 $\Delta(z)$ decays with increase in redshift as expected.}
\label{Fig(4)}
\end{figure}

 We consider \cite{inagaki2020gravitational} 
\begin{equation}
{\displaystyle{
 f_{2}(G)=-M_{pl}^{2}\Lambda \left(1-\alpha\exp^{-\frac{G}{G0}}\right)}},
\end{equation}
we assume $M_{pl}^{2}\Lambda$ is normalised to $1$, $\alpha =1$ and for $1-\exp^{(-\frac{G}{G0})}\rightarrow -G$, then $f(G)=G$ ( Eq. \ref{eq50} for dust dominated universe and 
Eq. \ref{eq52} for radiation dominated universe). Using the redshift transformation scheme and inserting Eq. \ref{eq63} through to Eq. \ref{eq65} into Eq. \ref{eq49}
 and Eq. \ref{eq50} and using $f_{2}(G)\equiv f$, numerical solutions for dust dominated universe are found and presented in Figure \ref{Fig(4)}
\newpage
For radiation dominated universe, we need to take into account the wavenumber dependency
\subsubsection{Short-wavelength solutions}
\begin{figure}[ht!]
 \includegraphics[width=120mm,height=100mm]{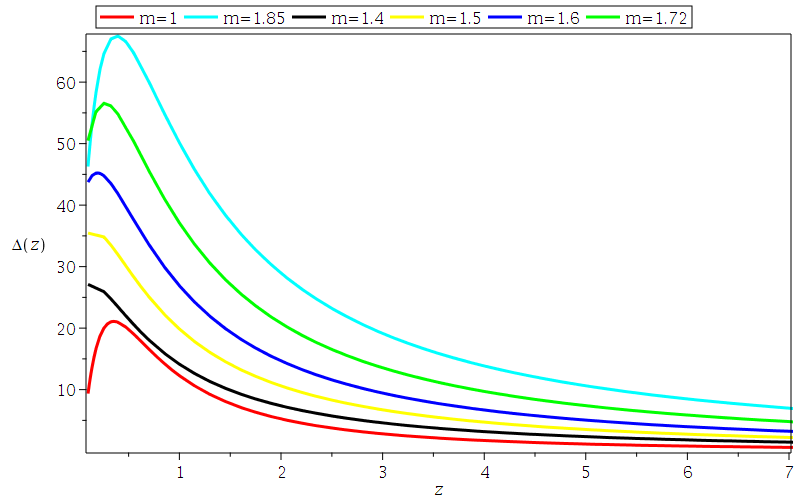}
 \caption[ Plot of energy-density perturbations $(\Delta(z))$ versus redshift $(z)$ of $f_{2}(G)$ model in radiation dominated universe, short wavelength modes.]
 {Plot of energy-density perturbations $(\Delta(z))$ versus redshift $(z)$ of Eq. \ref{eq51} and Eq. \ref{eq52}
 for $f_{2}(G)$ model in radiation dominated universe, short-wavelength limit. 
The red line corresponds to $f(G)=G$. The assumption 
that $\Delta^{(k)}_{(d)0}=10^{-5}$ and $\Delta'^{(k)}_{(d)0}=0$ was made and $\frac{k^{2}}{a^{2}H^{2}}\simeq\frac{16\pi^{2}}{3\lambda^{2}(1+z)^4}$
was considered.
 $\Delta(z)$ decays with increase in redshift as expected.} 
\label{Fig(5)}
\end{figure}

 Here we assume $\frac{k^{2}}{a^{2}}\gg 1$. In this regime, we consider that $\frac{k^{2}}{a^{2}H^{2}}\simeq\frac{16\pi^{2}}{3\lambda^{2}(1+z)^4}$ and 
 using the redshift transformation scheme and inserting Eq. \ref{eq66} through to Eq. \ref{eq68} into Eq. \ref{eq51} and Eq. \ref{eq52}
 and using $f_{2}(G)\equiv f$, numerical solutions for Gauss-Bonnet field-radiation dominated system short wavelength modes are 
   found and presented in Figure \ref{Fig(5)}.
\newpage
\subsubsection{ Long-wavelength solutions}
\begin{figure}[ht!]
 \includegraphics[width=120mm,height=100mm]{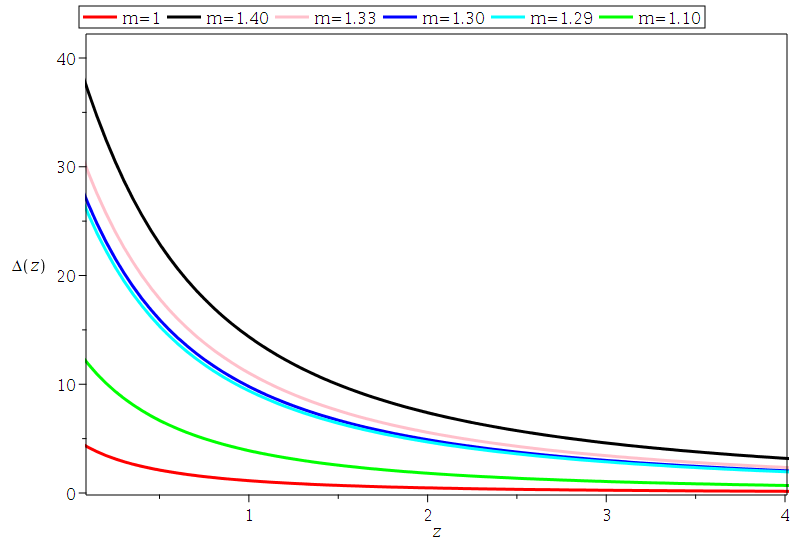}
 \caption[Plot of energy-density perturbations $(\Delta(z))$ versus redshift $(z)$ of $f_{2}(G)$ model in radiation dominated universe, long wavelength modes.]
 {
Plot of energy-density perturbations $(\Delta(z))$ versus redshift $(z)$ of Eq. \ref{eq51} and Eq. \ref{eq52} for  
$f_{2}(G)$ model in radiation dominated universe, long-wavelength limit. 
The red line corresponds to $f(G)=G$. The assumption 
that $\Delta^{(k)}_{(d)0}=10^{-5}$ and $\Delta'^{(k)}_{(d)0}=0$ was made and $\frac{k^{2}}{a^{2}H^{2}}\simeq\frac{16\pi^{2}}{3\lambda^{2}(1+z)^4}\simeq0$ was considered.
 $\Delta(z)$ decays with increase in redshift as expected.}
 \label{Fig(6)}
\end{figure}

 For the long wavelength regime we assume $\frac{k^{2}}{a^{2}}\ll 1$, which allow us to drop out $k$- dependency  ($\frac{k^{2}}{a^{2}H^{2}}\simeq\frac{16\pi^{2}}{3\lambda^{2}(1+z)^4}\simeq0$) 
and using the redshift transformation scheme and inserting Eq. \ref{eq66}
   through to Eq. \ref{eq68} into Eq. \ref{eq51} and Eq. \ref{eq52} and using $f_{2}(G)\equiv f$, numerical solutions for Gauss-Bonnet field-radiation dominated system long wavelength modes are 
   found and presented in Figure \ref{Fig(6)}.

\subsection{The logarithmic $f(G)$ model}
\begin{figure}[ht!]
 \includegraphics[width=120mm,height=100mm]{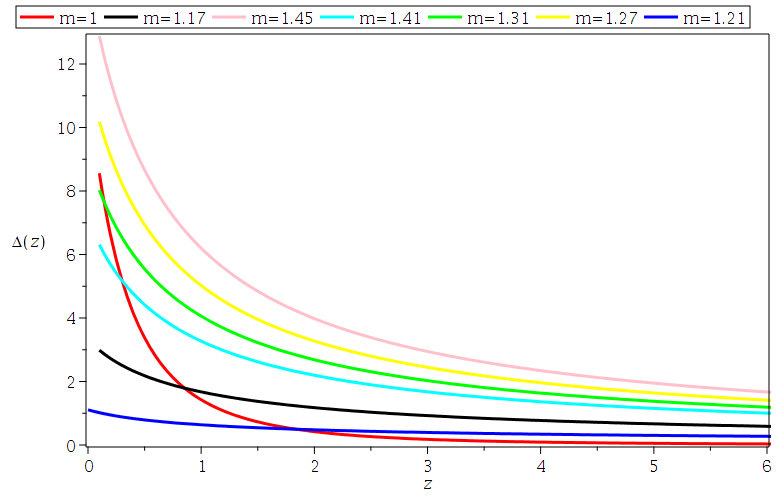}
 \caption[
  Plot of energy-density perturbations $(\Delta(z))$ versus redshift $(z)$ of $f_{2}(G)$ model in a dust dominated universe.]
 {Plot of energy-density perturbations $(\Delta(z))$ versus redshift $(z)$ of Eq. \ref{eq49} and Eq. \ref{eq50}
for $f_{3}(G)$ model in a dust dominated universe.
The red line corresponds to $f(G)=G$. The assumptions 
that $\Delta^{(k)}_{(d)0}=10^{-5}$ and $\Delta'^{(k)}_{(d)0}=0$ were made and 
 $\Delta(z)$ decays with increase in redshift as expected.}
\label{Fig(7)}
\end{figure}

 We consider \cite{zhou2009cosmological}
\begin{equation}
 f_{3}(G)=\ln(\frac{\alpha G}{G0}),
\end{equation}
 for the case $\alpha =1$ and for $\ln(\frac{G}{G0})\rightarrow G$, then $f(G)=G$ ( Eq. \ref{eq50} for dust dominated universe and 
Eq. \ref{eq52} for radiation dominated universe). For different $f(G)$ models, see for example \cite{shamir2020stellar, bamba2010finite, sharif2019ghost, bahamonde2020exact, 
nojiri2008inflation, de2009vacuum, garcia2011energy, odintsov2019dynamics, cognola2006dark, shamir2016dark, lee2020viable}.\\ \\
Using the redshift transformation scheme and inserting Eq. \ref{eq63} through to Eq. \ref{eq65} into Eq. \ref{eq49}
 and Eq. \ref{eq50} and using $f_{3}(G)\equiv f$, numerical solutions for dust dominated universe are found and presented in Figure \ref{Fig(7)}.
\newpage
For radiation dominated universe, we need to take into account the wavenumber dependency.
\subsubsection{Short-wavelength solutions}
\begin{figure}[ht!]
 \includegraphics[width=120mm,height=100mm]{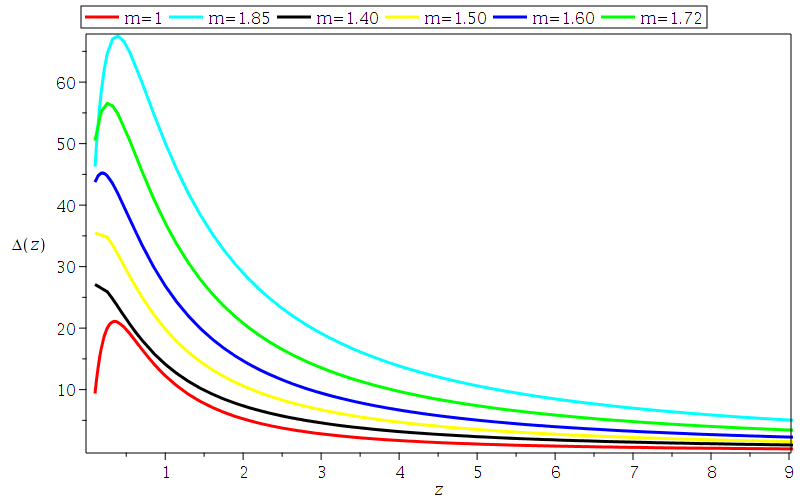}
 \caption[ Plot of energy-density perturbations $(\Delta(z))$ versus redshift $(z)$ of $f_{2}(G)$ model in radiation dominated universe, short wavelength modes.]
 {Plot of energy-density perturbations $(\Delta(z))$ versus redshift $(z)$ of Eq. \ref{eq51} and Eq. \ref{eq52}
 for $f_{3}(G)$ model in radiation dominated universe, short-wavelength limits. 
The red line corresponds to $f(G)=G$. The assumption 
that $\Delta^{(k)}_{(d)0}=10^{-5}$ and $\Delta'^{(k)}_{(d)0}=0$ was made and $\frac{k^{2}}{a^{2}H^{2}}\simeq\frac{16\pi^{2}}{3\lambda^{2}(1+z)^4}$ was considered.
 $\Delta(z)$ decays with increase in redshift as expected.}
\label{Fig(8)}
\end{figure}

 We assume $\frac{k^{2}}{a^{2}}\gg 1$, for which $\frac{k^{2}}{a^{2}H^{2}}\simeq\frac{16\pi^{2}}{3\lambda^{2}(1+z)^4}$. \\ \\ Using the 
 redshift transformation scheme and inserting Eq. \ref{eq66} through to Eq. \ref{eq68} into Eq. \ref{eq51} and Eq. \ref{eq52} and 
 using $f_{3}(G)\equiv f$, numerical solutions for Gauss-Bonnet field-radiation dominated system, short wavelength modes are 
   found and presented in Figure \ref{Fig(8)}.
\newpage
\subsubsection{ Long-wavelength solutions}
\begin{figure}[ht!]
 \includegraphics[width=120mm,height=100mm]{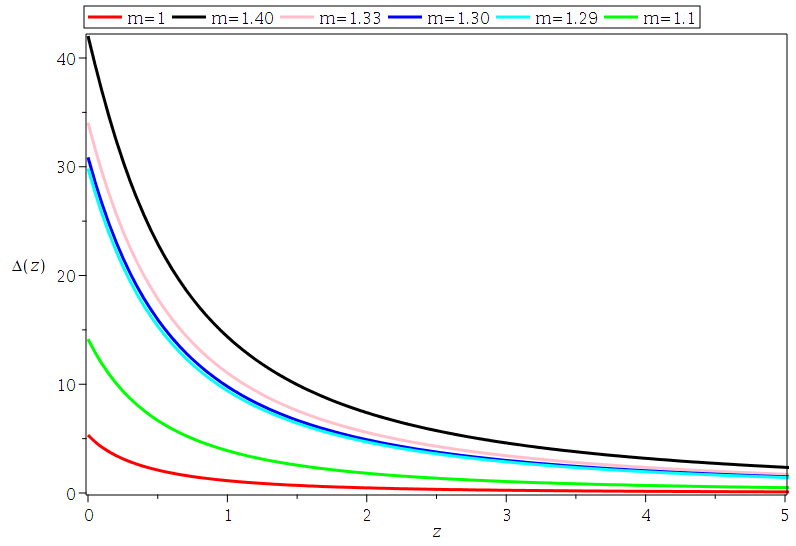}
 \caption[Plot of energy-density perturbations $(\Delta(z))$ versus redshift $(z)$ of $f_{2}(G)$ model in radiation dominated universe, long wavelength modes.]
 {Plot of energy-density perturbations $(\Delta(z))$ versus redshift $(z)$ of Eq. \ref{eq51} and Eq. \ref{eq52} for  
$f_{3}(G)$ model in radiation dominated universe, long-wavelength limits.
The red line corresponds to $f(G)=G$. The assumption 
that $\Delta^{(k)}_{(d)0}=10^{-5}$ and $\Delta'^{(k)}_{(d)0}=0$ was made and $\frac{k^{2}}{a^{2}H^{2}}\simeq\frac{16\pi^{2}}{3\lambda^{2}(1+z)^4}\simeq0$ was considered.
 $\Delta(z)$ decays with increase in redshift as expected.}
 \label{Fig(9)}
\end{figure}

We assume $\frac{k^{2}}{a^{2}}\ll 1$, and drop out $k$- dependency  ($\frac{k^{2}}{a^{2}H^{2}}\simeq\frac{16\pi^{2}}{3\lambda^{2}(1+z)^4}\simeq0$).\\ \\ 
 Using the redshift transformation scheme, inserting Eq. \ref{eq66}
   through to Eq. \ref{eq68} into Eq. \ref{eq51} and Eq. \ref{eq52} and using $f_{3}(G)\equiv f$, numerical solutions for Gauss-Bonnet field-radiation
   dominated system long wavelength modes are found and presented in Figure \ref{Fig(9)}.
\section{Discussions}
We studied cosmological perturbations in modified Gauss-Bonnet gravity. Using the $1+3$ covariant approach, we derived the linear 
covariant perturbations of a flat FRW spacetime background. We defined new gradient variables and derived their evolution equations, where after 
applying the spherical harmonic decomposition method and quasi-static approximation we get new equation, Eq.\ref{eq46} for energy-density
perturbations analysis in $f(G)$ gravity theory. After appying the redshift transformation , we focused on  the Gauss-Bonnet field-dust and Gauss-Bonnet field-radiation 
systems. We considered three different $f(G)$ models: the exponential, the logarithmic and the trigonometric.\\ \\ 
For Gauss-Bonnet field-dust system, the
 numerical solutions presented in Fig. \ref{Fig(1)} (trigonometric model), Fig. \ref{Fig(4)} (Exponential model) and Fig. \ref{Fig(7)} (Logarithmic model) show
 that the energy-density perturbations decay with increase in redshift for all the three $f(G)$ models. 
 The numerical solutions presented in these figures looks similar to those existing in the literatures for $f(R)$ and $f(T)$ gravity theories. If one is interested in 
  how matter perturbations behave in modified theories of gravity, see the work done in \cite{Ananda2009, Sahlu2020}.\\ \\
For Gauss-Bonnet field-radiation system, we considered short wavelength and long wavelength limits.
In the short-wavelength limit, we assumed that $k^{2}$ is much larger than the other terms.
  The numerical solutions presented in Fig. \ref{Fig(2)} (Trigonometric model), Fig. \ref{Fig(5)} (Exponential model) and Fig. \ref{Fig(8)} (Logarithmic model) 
  show that the energy-density perturbations ($\Delta(z)$) decay with increasing in redshift for all the three models and 
  the $\Delta(z)$ oscillates with decreasing amplitude for the trigonometric $f(G)$ model ( Fig. \ref{Fig(2)}). 
  The similar findings can be found in the work done in  \cite{ellis2012relativistic} for GR case  
and in the work done in \cite{abebe2013large} for $f(R)$ gravity theory.\\
In the long-wavelength limit, during numerical computation, we assumed that  $k^{2}$ is smaller enough compared to other terms, 
  therefore $k^{2}\approx 0$. The  numerical solutions presented in Fig. \ref{Fig(3)} (Trigonometric model), Fig. \ref{Fig(6)} (Exponential model) and Fig. \ref{Fig(9)} (Logarithmic model) show 
 that the energy-density perturbations decay with increase in redshift for all the three $f(G)$ models. The results for $f(R)$ and $f(T)$ gravity theories presented in  \cite{Ananda2009, Sahlu2020, abebe2013large} agree with our findings.\\ \\
  Some of the specific highlights of this work are as follows:
  in the trigonometric $f(G)$ model, we have shown the ranges of $m$ for which the Energy-density perturbations ($\Delta(z)$) oscillate or decay in both Gauss-Bonnet field-dust and 
  Gauss-Bonnet field-radiation systems. For example, in dust perturbations, there are no oscillating behaviors observed for $0.75 \leq m\leq 1$ while the $\Delta(z)$
   decay in this range. In radiation perturbations, $\Delta(z)$ depict oscillating behaviors in the short-wavelength limit for $1\leq m\leq 2$, and 
   $\Delta(z)$ decay for $1\leq m\leq 1.5$ in long-wavelength limit. In the exponential $f(G)$ model, we have shown that $\Delta(z)$ decay monotonically in the dust-dominated 
   perturbations for $1\leq m \leq 1.2$. For the radiation-dominated perturbations, there is no significant oscillating behavior observed in the 
    short-wavelength limit for $1\leq m \leq 2$ and the modes decay monotonically for the long-wavelength regime for $1\leq m\leq 1.5$. In the 
     logarithmic $f(G)$ model, in the dust-dominated perturbations, $\Delta(z)$ do not depict oscillating behavior but decay monotonically for $1\leq m\leq1.5$.
      In radiation-dominated perturbations, $\Delta(z)$ do not present oscillating behavior in short-wavelength limit for $1\leq m\leq 2$ as well as 
      in long-wavelength limit for $1\leq m \leq 1.5$ but in these ranges, the energy-density perturbations decay with increase in redshift.
      The choice of the values of $m$ follows the work done in \cite{Ananda2009, abebe2013large}.
  \section{Conclusions }\label{conclusion}
This work presents a detailed analysis of cosmological perturbations in modified Gauss-Bonnet gravity theory using a $1+3$
 covariant formalism. We defined vector and scalar gradient variables and derived the corresponding evolution equations. Using spherical harmonic decomposition method,
 we  were able to obtain the ordinary differential equations (ODEs) manageable for the analysis. These ODEs were then transformed to be redshift dependent. The obtained equations 
 for the matter energy-density and for the Gauss-Bonnet energy density were coupled, then decoupled using quasi-static approximation to make equations manageable.\\ \\
 We considered three different viable models $f_{1}(G)=\frac{\alpha G}{\sqrt{G0}}\arctan(\frac{G}{G0})-\alpha \lambda \sqrt{G0}$, $f_{2}(G)=-M_{pl}^{2}\Lambda (1-\alpha\exp^{-\frac{G}{G0}})$
 and $f_{3}(G)=\ln(\alpha \frac{G}{G0})$ and we found that 
 for Gauss-Bonnet field-dust system, the matter energy-density perturbations $\Delta(z)$ decay with increase in 
 redshift for all of the three $f(G)$ gravity models.
 In the case of Gauss-Bonnet field-radiation system, we considered short-wavelength and long-wavelength modes 
 and found that the $\Delta(z)$ for the long-wavelength modes decays  with increase in redshift whereas  
 for the short-wavelength modes, $\Delta(z)$ decays with increase in redshift and oscillates with a decreasing in amplitude 
  for the trigonometric $f(G)$ model.
   We conclude that for all of the three considered $f(G)$ models, 
 the  model parameters can be constrained using observational data and can be fit to the currently known features 
 of the large scale structure matter power spectrum  in modified gravity theories.
 Analysis of energy-density perturbations for a multi-fluid system is left to 
 the future research.
\section*{Acknowledgements} 
  Albert Munyeshyaka gratefully acknowledges financial support from the Swedish International Development Cooperation Agency (SIDA) through the International Science Programme (ISP) to University of Rwanda
  through the East African Astrophysics Research Network (EAARN), project number AFRO:05.\\
  JN gratefully acknowledges financial support from the Swedish International Development Cooperation Agency (SIDA) through  ISP to the University of Rwanda through Rwanda Astrophysics, Space and Climate Science Research Group (RASCSRG), project number RWA:01.
 \bibliographystyle{unsrt}
\bibliography{Albert}
 
 \section*{Appendix}
 \begin{multline*}
 A=\{\frac{1}{\theta^{2}f''}(\frac{27}{32\theta^{2}}+\frac{3}{2}f')
 +\frac{f'''}{\theta^{2}f''}(-\frac{27}{32}\frac{G}{\theta^{2}}-\frac{\theta^{2}}{4}
  -\frac{3(1-3w)}{4}\rho_{m}+\frac{3}{2}f\\
  -\frac{3}{2}Gf'+\dot{G}^{2}\theta^{2}f''')
  -\frac{3}{2\theta}f'+\frac{3}{2\theta^{2}}G
  -\frac{9}{4\theta}^{3}\dot{G}-\frac{\dot{G}^{2}f^{iv}}{f''}\}.
\end{multline*}
\begin{multline*}
 B_{1}=\frac{1}{\theta^{4}}\{\frac{27}{64A}\rho_{d}(Gf'''-f'')\}+\frac{1}{\theta^{2}}\{\frac{3}{4A}\rho_{d}(\frac{f''f''}{2}
 -Gf''-\frac{3}{2}G\dot{G}f'''f''\\
 +\frac{\rho_{d}}{2}f'''-f'''f+f'''f'-f''f')\}+\frac{1}{\theta}(\frac{3}{4A}\rho_{d}f''f')+\theta(\frac{1}{2A}\rho_{d}\dot{G}f'''f'')
 \\+\rho_{d}
 +\frac{\rho_{d}}{A}\{-\dot{G}f'''f''-\frac{1}{2}\ddot{G}f'''f''+f'''-\frac{1}{2}\dot{G}^{2}f'''f'''\}.
\end{multline*}
Using
\begin{multline*}
 B_{11}= \frac{1}{\theta^{7}}(\frac{243}{128Af''})+\frac{1}{\theta^{6}}\{\frac{81}{16A}G\dot{G}\frac{f'''}{f''}
 +\frac{243}{64}\dot{G}G^{2}\frac{f'''}{f''}-\frac{243}{128}G\dot{G}\}\\
   +\frac{1}{\theta^{5}}\{\frac{27}{8A}(  \frac{f'}{f''}+\frac{\rho_{d}}{4f''}
   -\frac{f}{2f''}+\frac{Gf'}{2f''})\\+
   \frac{27}{8A}(-\frac{1}{2}+G) +\frac{27}{16A}(-\rho_{d}\frac{G}{2}\frac{f'''}{f''f''}+G\frac{ff'''}{f''f''}-G^{2}\frac{f'f'''}{f''f''}\\
   -2\frac{ff'''}{f''f''}
   -2G\frac{f'f'''}{f''f''})-\frac{81}{8A}G^{2}\dot{G^{2}}f'''\},
   \end{multline*}
   \begin{multline*}
    B_{12}=\frac{1}{\theta^{4}}\{\frac{9}{2A}(\rho_{d}G\dot{G}\frac{f'''}{f''}-G\dot{G}f\frac{f'''}{f''}
   G^{2}\dot{G}f'\frac{f'''}{f''})\\
   +\frac{27}{4}(G\dot{G}\frac{f'''}{2f''}-G\dot{G}f\frac{f'''}{f''}-G^{2}\dot{G}f'\frac{f'''}{f''})
   +\frac{27}{4A}(\frac{f''}{2}-G^{2}\dot{G}f'')
   -\frac{27}{4}G\dot{G}f-\frac{27}{8A}f''\} \\
   +\frac{1}{\theta^{3}}\{  \frac{3}{A}(\frac{\rho_{d}f'}{2f''}-\frac{ff'}{f''}+\frac{Gf'f'}{f''})
   +\frac{3}{2A}(-\frac{\rho_{d}}{2}+f-f'+\rho_{d}G-2Gf+2Gf'+\frac{3}{8}\frac{f'''}{f''f''}\\
   -\frac{\rho^{2}_{d}}{2}\frac{f'''}{f''f''}-2\rho_{d}G\frac{f'f'''}{f''f''}
   -2\frac{fff'''}{f''f''}
   -2G^{2}\frac{f'f'f'''}{f''f''}+\frac{3}{2}\dot{G}^{2}\frac{f'''f'''}{f''f''})+\frac{27}{4A}G\dot{G}f'f''\\ +\frac{9}{2A}\dot{G}\frac{f'''}{f''}-
   \ddot{G}\frac{f'''}{2f''}\},
   \end{multline*}
   \begin{multline*}
   B_{13}=\frac{1}{\theta^{2}}\{ \frac{3}{A}(-\frac{\rho_{d}}{2}f'+ff'-f'f')
   +\frac{9}{4}-3G\dot{G}f''+\frac{3}{16}\dot{G}\\+\frac{15}{16}G\dot{G}\frac{f'''}{f''}
   -\frac{9}{2}G\dot{G}^{3}\frac{f'''f'''}{f''}-\frac{9}{2A}(G\dot{G}\ddot{G}f'''+\dot{G}\frac{f'''}{2f''})\}
   +\frac{1}{\theta}\{\frac{1}{A}(2\rho_{d}\dot{G}\frac{f'''}{f''}
   \ddot{G}\frac{f'''}{f''}\\-4\dot{G}\frac{ff'''}{f''}
   -2\ddot{G}\frac{ff'''}{f''}+4G\dot{G}\frac{f'f'''}{f''}+2G\dot{G}\frac{f'f'''}{f''}-\rho_{d}\frac{f'''}{4f''f''} 
   +\frac{ff'''}{2f''f''}-G\frac{f'f'''}{2f''f''}\\+\rho_{d}\dot{G}^{2}\frac{f'''f'''}{f''f''}
   -2\dot{G}^{2}\frac{ff'''f'''}{f''f''}+2G\dot{G}^{2}\frac{f'f'''f'''}{f''f''})
   +\frac{5}{A}G\dot{G}^{2}f'''+\rho_{d}-\frac{f}{2}+G\frac{f'}{2}\},
   \end{multline*}
   \begin{multline*}
   B_{14}=\theta\{ -\frac{2}{3}
   -\frac{4}{3}\dot{G}^{2}f'''-\frac{4}{3}\ddot{G}f'' -\dot{G}\frac{f''f'}{3A} \}+\theta^{2}\{\frac{2}{3A}(\frac{2}{3}G\dot{G}f'''+\dot{G}\ddot{G}f''')+2\dot{G}^{3}\frac{f'''f'''}{f''}\\
   -\dot{G}\frac{f'''}{18f''}+
   \frac{8}{9}\dot{G}f''\}
   -\theta^{3}(\frac{2}{9A}f''')-\rho_{d}\dot{G}\frac{f'''}{6f''}+\dot{G}\frac{ff'''}{3f''}+G\dot{G}\frac{f'''}{3f''}\\
   +\frac{1}{A}(G\dot{G}\frac{f''}{3}-\dot{G}\frac{f''}{6}
   -\rho_{d}\dot{G}\frac{f'''}{f''}+2\dot{G}\frac{ff'''}{f''}-2G\dot{G}\frac{f'f'''}{f''})+\dot{G}\frac{f}{3}\},
\end{multline*}
we have
\begin{equation*}
 B_{2}=B_{11}+B_{12}+B_{13}+B_{14}.
\end{equation*}
Let 
\begin{multline*}
 C_{1}=-\frac{k^{2}}{3a^{2}}+\frac{4\rho_{r}}{3}-\frac{4\theta^{3}}{9} \dot{G}f''
 +\frac{2\theta^{2}}{9}-\frac{3}{4\theta}+\frac{1}{6}f
 -\frac{G}{6}f'
 +\frac{4\theta^{2}\dot{G}^{2}}{9}f'''
 +\frac{4\theta^{2}\ddot{G}}{9}f''+\frac{G\dot{G}}{2\theta}f''\\ 
 \frac{1}{24A}\theta\dot{G}f''+\frac{9}{32\theta^{3}A}G\dot{G}f''+\frac{1}{4A}\dot{G}^{2}f'''+\frac{1}{8A}\dot{G}\ddot{G}
 +\frac{1}{12A}\theta^{2}\dot{G}f'f''\\+\frac{9}{16\theta^{2}A}G\dot{G}f'f''
 +\frac{1}{2A}\theta\dot{G}^{2}f'f'''+\frac{1}{4A}\theta \dot{G}\ddot{G}f'f''
 -\frac{1}{12A}\theta G\dot{G}f''-\frac{9}{16\theta^{3}A}G\dot{G},
\end{multline*}
\begin{multline*}
 C_{2}=-\frac{1}{2A}G\dot{G}^{2}f'''-\frac{1}{4A}G\dot{G}\ddot{G}f''
 +\frac{1}{18A}\theta^{4}\dot{G}^{2}f'''
 +\frac{3}{8A}G\dot{G}^{2}f'''+\frac{1}{3A}\theta^{2}\dot{G}^{3}\frac{f'''f'''}{f''}\\
 +\frac{1}{6A}\theta^{3}\dot{G}^{2}\ddot{G}f'''-\frac{1}{8A}G\dot{G}^{2}f'''
 -\frac{27}{32\theta^{4}A}G^{2}\dot{G}^{2}f'''
 -\frac{3}{4\theta A}G\dot{G}^{3}\frac{f'''f'''}{f''}-\frac{3}{8\theta A}G\dot{G}^{2}\ddot{G}f'''\\
 -\frac{1}{9A}\theta^{3}\dot{G}^{2}f'''-\frac{3}{4\theta A}G\dot{G}^{2}f'''-\frac{2}{3A}\theta^{2}\dot{G}^{3}\frac{f'''f'''}{f''}-\frac{1}{3A}\theta^{2}\dot{G}^{2}\ddot{G}f'''
 \\-\frac{1}{18A}\theta^{3}\dot{G}\ddot{G}f'''-\frac{3}{8A}\theta G\dot{G}\ddot{G}f'''-\frac{1}{3A}\theta^{2}\dot{G}^{2}\ddot{G}\frac{f'''f'''}{f''}
 -\frac{1}{6A}\theta^{2}\dot{G}\ddot{G}^{2}f'''+\frac{3}{64\theta A}G\dot{G}\frac{f'''}{f''}\\
 +\frac{81}{256\theta^{5}A}G^{2}\dot{G}\frac{f'''}{f''}+\frac{9}{32\theta^{2}A}G\dot{G}^{2}\frac{f'''f'''}{f''f''}
 +\frac{9}{64\theta^{2}A}G\dot{G}\ddot{G}\frac{f'''}{f''}+\frac{1}{192A}\theta^{3}\dot{G}\frac{f'''}{f''}+\frac{3}{32\theta A}G\dot{G}\frac{f'''}{f''}\\
 +\frac{1}{12A}\theta^{2}\dot{G}^{2}\frac{f'''f'''}{f''f''}
 +\frac{1}{24A}\theta^{2}\dot{G}\ddot{G}\frac{f'''}{f''}-\frac{1}{12A}\theta\dot{G}\frac{ff'''}{f''}-\frac{9}{16\theta^{3}A}G\dot{G}\frac{ff'''}{f''}-
 \frac{1}{2A}\dot{G}^{2}\frac{f'''ff'''}{f''f''}\\-\frac{1}{4A}\dot{G}\ddot{G}\frac{ff'''}{f''}
 +\frac{1}{12A}\theta G\dot{G}\frac{f'f'''}{f''}+\frac{9}{16\theta^{3}A}G^{2}\dot{G}\frac{f'f'''}{f''}
 +\frac{1}{2A}G\dot{G}^{2}\frac{f'''f'f'''}{f''f''}+\frac{1}{4A}G\dot{G}\ddot{G}\frac{f'f'''}{f''},
\end{multline*}
\begin{multline*}
 C_{3}=
 -\frac{1}{18A}\theta^{3}\dot{G}^{3}\frac{f'''f'''}{f''}
 -\frac{3}{8\theta A}G\dot{G}^{3}\frac{f'''f'''}{f''}-\frac{1}{3A}\theta^{2}\dot{G}^{4}\frac{f'''f'''f'''}{f''f''}-\frac{1}{6A}\theta^{2}\dot{G}^{3}\ddot{G}\frac{f'''f'''}{f''}
 -\frac{1}{12A}\theta \dot{G}f'\\-\frac{9}{16\theta^{3}A}G\dot{G}f'
 -\frac{1}{2A}\dot{G}^{2}\frac{f'f'''}{f''}-\frac{1}{4A}\dot{G}\ddot{G}f'-\frac{9}{64\theta A}\dot{G}
 -\frac{81}{256\theta^{5}A}G\dot{G}-\frac{9}{32\theta^{2}A}\dot{G}^{2}\frac{f'''}{f''}-\frac{9}{64\theta^{2}A}\dot{G}\ddot{G}\\
   +\frac{9}{16A\theta^{4}}-\frac{f}{2A\theta^{2}}+\frac{Gf'}{2A\theta^{2}}+\frac{9f'}{8A\theta^{3}}-\frac{ff'}{A\theta}+\frac{Gf'f'}{A\theta^{2}}-\frac{9G}{8A\theta^{4}}
    +\frac{Gf}{A\theta^{2}}-\frac{G^{2}f'}{A\theta^{2}}+\frac{3}{4A\theta}\dot{G}\frac{f'''}{f''}-\frac{2\theta}{3A}\dot{G}\frac{ff'''}{f''}\\
    +\frac{2\theta}{3A}G\frac{f'f'''}{f''}
   -\frac{27}{16A\theta^{5}}G\dot{G}\frac{f'''}{f''}+ \frac{3}{2A\theta^{3}}G\dot{G}\frac{ff'''}{f''}-\frac{3}{2A\theta^{3}}G^{2}\dot{G}\frac{f'f'''}{f''}
   -\frac{3}{2A\theta^{2}}\dot{G}\frac{f'''}{f''}+\frac{4}{3A}\dot{G}\frac{ff'''}{f''},
\end{multline*}
\begin{multline*}
 C_{4}=-\frac{4}{3A}G\dot{G}\frac{f'f'''}{f''}-\frac{3}{4A\theta^{2}}\ddot{G}\frac{f'''}{f''}
    +\frac{2}{3A}\ddot{G}\frac{ff'''}{f''}-\frac{2}{3A}G\ddot{G}\frac{f'f'''}{f''}+\frac{81}{128A\theta^{6}}G\frac{f'''}{f''f''}-\frac{9}{16A\theta^{4}}G\frac{ff'''}{f''f''}\\
    +\frac{9}{16A\theta^{4}}G^{2}\frac{f'f'''}{f''f''}+\frac{3}{16A\theta^{2}}\frac{f'''}{f''}-\frac{1}{6A}\frac{ff'''}{f''f''}+\frac{1}{6A}G\frac{f'f'''}{f''f''}
     -\frac{9}{8A\theta^{4}}\frac{ff'''}{f''f''}+\frac{1}{A\theta^{2}}\frac{fff'''}{f''f''}\\
     -\frac{1}{A\theta^{2}}G\frac{ff'f'''}{f''f''}+\frac{9}{8A\theta^{4}}G\frac{f'f'''}{f''f''}
     -\frac{1}{A\theta^{2}}G\frac{ff'f'''}{f''f''}+\frac{1}{A\theta^{2}}G^{2}\frac{f'f'f'''}{f''f''}-\frac{3}{4A\theta^{2}}\dot{G}^{2}\frac{f'''f'''}{f''f''}
     \\+ \frac{2}{3A}\dot{G}^{2}\frac{ff'''f'''}{f''f''}
     -\frac{2}{3A}G\dot{G}^{2}\frac{f'f'''f'''}{f''f''}-\frac{9}{8A\theta^{4}}\frac{f'}{f''}+\frac{1}{A\theta^{2}}\frac{ff'}{f''}
     -\frac{1}{A\theta^{2}}G\frac{f'f'}{f''}-\frac{81}{128A\theta^{6}}\frac{1}{f''},
\end{multline*}
\begin{multline*}
 C_{5}=\frac{9}{16A\theta^{4}}\frac{f}{f''}-\frac{9}{16A\theta^{4}}G\frac{f'}{f''} 
      +\frac{1}{18A}\theta\dot{G}f''-\frac{9}{8A\theta^{3}}G\dot{G}f''+\frac{1}{9A}\theta^{2}\dot{G}f'f''-\frac{9}{4A\theta^{2}}G\dot{G}f'f''\\
 -\frac{1}{9A}\theta G\dot{G}f''
      +\frac{9}{4A\theta^{3}}G^{2}\dot{G}f'' +\frac{2}{27A}\theta^{4}\dot{G}^{2}f'''
      -\frac{3}{2A}G\dot{G}^{2}f'''-\frac{1}{6A}G\dot{G}^{2}f'''+\frac{27}{8A\theta^{4}}G^{2}\dot{G}^{2}f''' \\
       -\frac{4}{27A}\theta^{3}\dot{G}^{2}f'''+\frac{3}{A\theta}G\dot{G}^{2}f'''-\frac{2}{27A}\theta^{3}\dot{G}\ddot{G}f'''+\frac{3}{2A\theta}G\dot{G}\ddot{G}f'''
   +\frac{1}{16A\theta}G\dot{G}\frac{f'''}{f''}-\frac{81}{64A\theta^{5}}G^{2}\dot{G}\frac{f'''}{f''},
\end{multline*}
\begin{multline*}
 C_{6}=\frac{1}{54A}\theta^{3}\dot{G}\frac{f'''}{f''}-\frac{3}{8A\theta}G\dot{G}\frac{f'''}{f''}
   -\frac{1}{9A}\theta \dot{G}\frac{ff'''}{f''}+\frac{9}{4A\theta^{3}}G\dot{G}\frac{ff'''}{f''}+\frac{1}{9A}\theta G\dot{G}\frac{f'f'''}{f''}\\
   -\frac{9}{4A \theta^{3}}G^{2}\dot{G} \frac{f'f'''}{f''}
     -\frac{2}{27A}\theta^{3}\dot{G}^{3}\frac{f'''f'''}{f''}+\frac{3}{2A\theta}G\dot{G}^{3}\frac{f'''f'''}{f''} -\frac{1}{9A}\theta \dot{G}f'+\frac{9}{4A\theta^{3}}G\dot{G}f'
     -\frac{1}{16A\theta}\dot{G}\\+\frac{81}{64A\theta^{5}}G\dot{G},
\end{multline*}
we can have
\begin{equation*}
 C=C_{1}+C_{2}+C_{3}+C_{4}+C_{5}+C_{6}.
\end{equation*}
Considering
\begin{multline*}
 D_{1}=-\frac{2\theta}{3}+\frac{\rho_{d}}{\theta}+\frac{9}{4\theta^{2}}-\frac{1}{2\theta}f
 +\frac{G}{2\theta}f'-\frac{4\theta\dot{G}^{2}}{3}f'''
 +\frac{8\theta^{2}\dot{G}}{9}f''+\frac{3G\dot{G}}{2\theta^{2}}f''\\-\frac{4\theta\ddot{G}}{3}f''-\frac{9G\dot{G}}{2\theta^{2}}f'' 
 -\frac{27}{16A\theta^{5}}+\frac{3}{2A\theta^{3}}f
 -\frac{3}{2A\theta^{3}}Gf'
 -\frac{27}{8A\theta^{4}}f'+\frac{3}{A\theta^{2}}ff'-\frac{3}{A\theta^{2}}Gf'f'+\frac{27}{8A\theta^{5}}G\\
 -\frac{3}{A\theta^{3}}Gf+\frac{3}{A\theta^{3}}G^{2}f'-\frac{9}{4A\theta^{2}}\dot{G}\frac{f'''}{f''}+\frac{2}{A}\dot{G}\frac{ff'''}{f''}
   -\frac{2}{A}G\dot{G}\frac{f'f'''}{f''}+\frac{81}{16A\theta^{6}}G\dot{G}\frac{f'''}{f''}-\frac{9}{2A\theta^{4}}G\dot{G}\frac{ff'''}{f''},
\end{multline*}
\begin{multline*}
 D_{2}=\frac{9}{2A\theta^{4}}G^{2}\dot{G}\frac{f'f'''}{f''} 
 +\frac{9}{2A\theta^{3}}\dot{G}\frac{f'''}{f''}-\frac{4}{A\theta}\dot{G}\frac{ff'''}{f''}+\frac{4}{A\theta}G\dot{G}\frac{f'f'''}{f''}
 +\frac{9}{4A\theta^{3}}\ddot{G}\frac{f'''}{f''}
 -\frac{2}{A\theta}\ddot{G}\frac{ff'''}{f''}\\
 +\frac{2}{A\theta}G\ddot{G}\frac{f'f'''}{f''}-\frac{243}{128A\theta^{7}}G\frac{f'''}{f''f''}+\frac{27}{16A\theta^{5}}G\frac{ff'''}{f''f''}
   -\frac{27}{16A\theta^{5}}G^{2}\frac{f'f'''}{f''f''}-\frac{9}{16A\theta^{3}}\frac{f'''}{f''f''}\\
 +\frac{1}{2A\theta}\frac{ff'''}{f''f''}-\frac{1}{2a\theta}G\frac{f'f'''}{f''f''}
   +\frac{27}{8A\theta^{5}}\frac{ff'''}{f''f''}
   -\frac{3}{A\theta^{3}}\frac{fff'''}{f''f''}+\frac{3}{A\theta^{3}}G\frac{f'f'''}{f''f''}-\frac{27}{8A\theta^{5}}G\frac{f'f'''}{f''f''},
\end{multline*}
\begin{multline*}
 D_{3}=\frac{3}{A\theta^{3}}G\frac{ff'f'''}{f''f''}
  -\frac{3}{A\theta^{3}}G^{2}\frac{f'f'f'''}{f''f''}+\frac{9}{4A\theta^{3}}\dot{G}^{2}\frac{f'''f'''}{f''f''}
    -\frac{2}{A\theta}\dot{G}^{2}\frac{ff'''f'''}{f''f''}+\frac{2}{A\theta}G\dot{G}^{2}\frac{f'f'''f'''}{f''f''}\\+\frac{27}{8A\theta^{5}}\frac{f'}{f''}
    -\frac{3}{A\theta^{3}}\frac{ff'}{f''}
    +\frac{3}{A\theta^{3}}G\frac{f'f'}{f''}+\frac{243}{128A\theta^{7}}\frac{1}{f''}-\frac{27}{16A\theta^{5}}\frac{f}{f''}
    +\frac{27}{16A\theta^{5}}G\frac{f'}{f''}-\frac{1}{6A}\dot{G}f''\\
    -\frac{1}{3A}\dot{G}\theta f'f''-\frac{1}{3A}G\dot{G}f''+\frac{27}{8A\theta^{4}}G\dot{G}f''
        +\frac{27}{4A\theta^{4}}G\dot{G}f'f''-\frac{27}{4A\theta}G^{2}\dot{G}f''+\frac{2}{9A}\theta^{3}\dot{G}^{2}f'''\\
        +\frac{1}{2A\theta}G\dot{G}^{2}f'''
    +\frac{4}{9A}\theta^{2}\dot{G}^{2}f'''+\frac{2}{9A}\theta^{2}\dot{G}\ddot{Gf}f'''+\frac{9}{2A\theta}G\dot{G}^{2}f''',
\end{multline*}
\begin{multline*}
 D_{4}=-\frac{81}{8A\theta^{5}}G^{2}\dot{G}^{2}f'''
    -\frac{9}{A\theta^{2}}G\dot{G}^{2}f'''-\frac{9}{2A\theta^{2}}G\dot{G}^{2}f'''+\frac{9}{2A\theta^{2}}G\dot{G}\ddot{G}f'''+\frac{1}{3}\dot{G}f'\\+\frac{3}{16\theta^{2}}\dot{G} 
      -\frac{27}{4\theta^{4}}G\dot{G}f'-\frac{729}{64\theta^{6}}G\dot{G}
      -\frac{3}{16A\theta^{2}}G\dot{G}\frac{f'''}{f''}-\frac{1}{18A}\theta^{2}\dot{G}\frac{f'''}{f''}
    +\frac{1}{3A}\dot{G}\frac{ff'''}{f''}\\-\frac{1}{3A}G\dot{G}\frac{f'f'''}{f''}+\frac{2}{9A}\theta^{2}\dot{G}^{3}\frac{f'''f'''}{f''}+\frac{243}{64A\theta^{6}}G^{2}\dot{G}\frac{f'''}{f''}
     +\frac{9}{8A\theta^{2}}G\dot{G}\frac{f'''}{f''}-\frac{27}{4A\theta^{4}}G\dot{G}\frac{ff'''}{f''}\\
     +\frac{27}{4A\theta^{4}}G^{2}\dot{G}\frac{f'f'''}{f''}
     -\frac{9}{2A\theta^{2}}G\dot{G}^{3}\frac{f'''f'''}{f''},
\end{multline*}
we have
\begin{equation*}
 D=D_{1}+D_{2}+D_{3}+D_{4}.
\end{equation*}
 
\end{document}